\documentstyle[11pt,aaspp4]{article}
\def\gax{\mathrel{\raise.3ex\hbox{$>$}\mkern-14mu\lower0.6ex\hbox{$\sim$}}}
\def\lax{\mathrel{\raise.3ex\hbox{$<$}\mkern-14mu\lower0.6ex\hbox{$\sim$}}}
\def\gtorder{\mathrel{\raise.3ex\hbox{$>$}\mkern-14mu
             \lower0.6ex\hbox{$\sim$}}}
\def\ltorder{\mathrel{\raise.3ex\hbox{$<$}\mkern-14mu
             \lower0.6ex\hbox{$\sim$}}}

\input psfig.sty

\begin{document}

\title{Quantitative Interpretation of Quasar Microlensing Light Curves}

\author{C.S. Kochanek}
\affil{Harvard-Smithsonian Center for Astrophysics, 60 Garden
        St., Cambridge, MA 02138}
\affil{email: ckochanek@cfa.harvard.edu}

\def\avgm{\langle M \rangle}
\def\avgmhat{\langle M/M_\odot \rangle}

\begin{abstract}
We develop a general method for analyzing the light curves of microlensed quasars and
apply it to the OGLE light curves of the four-image lens Q2237+0305.  We simultaneously
estimate the effective source velocity, the average stellar mass, the stellar mass function, 
and the size and structure of the quasar accretion disk.  The light curves imply
an effective source plane velocity of 
$10200~\hbox{km/s} \ltorder v_e h \avgmhat^{-1/2} \ltorder 39600$~km/s
(68\% confidence).  Given an independent estimate for the source velocity, 
found by combining 
estimates for the peculiar velocity of the lens galaxy with its measured stellar 
velocity dispersion, we obtain a mean stellar mass of
$\langle M \rangle \simeq 0.037 h^2 M_\odot$ 
($0.0059 h^2 M_\odot \ltorder \avgm \ltorder 0.20 h^2 M_\odot$).  We were
unable to distinguish a Salpeter mass function from one in which all stars had
the same mass, but we do find a strong lower bound of $\kappa_*/\kappa \gtorder 0.5$
on the fraction of the surface mass density represented by the microlenses.
Our models favor a standard thin accretion disk model
as the source structure over a simple Gaussian source.  For a face-on, thin disk 
radiating as a black body with temperature profile $T_s \propto R^{-3/4}$, 
the radius $r_s$ where the temperature matches the filter pass band (2000{\AA} or
$T_s(r_s)\simeq 7\times 10^4$~K) is 
$ 1.4\times 10^{15} h^{-1} \hbox{cm} \ltorder r_s \ltorder 4.5 \times 10^{15} h^{-1}\hbox{cm}$.
The flux predicted by the disk model agrees with the observed flux of the
quasar, so non-thermal or optically thin emission processes are not required.
From the disk structure we estimate a black hole mass of 
$M_{BH}\simeq 1.1_{-0.7}^{+1.4} \times 10^9 h^{-3/2}\eta_{0.1}^{1/2}(L/L_E)^{-1/2} M_\odot$, 
consistent with the mass estimated under the assumption that the quasar is
radiating at the Eddington luminosity ($L/L_E=1$).
\end{abstract}


\section{Introduction \label{sec:intro} }

The term ``microlensing'' describes the flux variations produced in a background
source by foreground stars in two very different regimes.  Today, astronomers are
most familiar with the local (Galactic) phenomenon, in which 
a star or binary produces a time variable magnification of a background star
(see the reviews by Paczynski~\cite{Paczynski96} or Mao~\cite{Mao01}).  
Because the physical distances are so short, the Galaxy is optically
thin to microlensing ($\tau \sim 10^{-6}$).  This leads to the disadvantage that
few background sources are lensed (one in $\tau^{-1}$ stars), and the advantage that the 
lens producing the variations is simple (one or two isolated stars).  In quasar
microlensing, by contrast, the existence of multiple quasar images requires a
microlensing optical depth near unity ($\tau \sim 1$) if the stars in the lens
galaxy are a significant fraction of the surface density (see the review
by Wambsganss~\cite{Wambsganss01}).  This 
regime has the advantage that all background sources are microlensed, but
the disadvantage that the lens is intrinsically complex, as it consists of a star 
field rather than a star.   

In either experiment, the light curve of
the background source provides a time history of the changes in the 
magnification created by the relative motions of the observer, the source,
and the lens.  At its simplest, these variations determine a time scale,
$\Delta t \propto M^{1/2} v_e^{-1} x^{1/2}(1-x)^{1/2}$, set by the mass $M$, 
the effective source velocity $v_e$, and the fractional distance $x$ of the lens
from the source.  These scalings are exact for Galactic microlensing events, and 
the stellar mass can be inferred only from the statistical properties of large samples 
(e.g. Alcock et al.~\cite{Alcock00}) or 
from events where special circumstances allow an independent determination of $v_e$ or $x$
(e.g. parallax effects, Grieger, Kayser \& Refsdal~\cite{Grieger86}, Gould~\cite{Gould92}).
For quasar microlensing, these same factors determine the typical time between
``events'' in which there is a significant change in the magnification, with the
advantage that the fractional distance $x$ is known from the redshifts,
leaving only a degeneracy between the mass and
velocity scales.  If the fundamental physics probed by the two regimes is the
same, why has the astronomical community devoted far more
observational resources to Galactic microlensing than to extragalactic microlensing?  

The first problem is that the time scales for quasar microlensing are roughly ten times
longer than for Galactic microlensing, because the larger length scales of the
extragalactic regime are only partly balanced by the larger velocity scales. 
As a result, ``events'' take 1--10 years rather than 0.1--1 years.  This
is no longer a viable argument for ignoring quasar microlensing.  There
are roughly 40 multiply-imaged quasars that could be monitored, with a total
of roughly 120 images, so that even if the ``event'' rate is only one per
image per decade, there are 10 quasar microlensing ``events'' occurring
each year.  Quasar microlensing requires less intensive monitoring because
of the longer time scales (once per week rather than once per day), so 
a total of roughly 2000 images/year is needed to monitor the available 
lens sample.  Even with the addition of more intensive monitoring during
events, this represents a small fraction of the effort in a large Galactic
microlensing survey.  

The second problem is that the quasar images are separated by only 
arcseconds, making it difficult to obtain the independent image fluxes.
Fortunately, many telescopes routinely produce sub-arcsecond resolution
images. When combined with difference imaging (e.g. Alard~\cite{Alard00})
to compensate for PSF variations with epoch and to remove the non-varying
components of the lens, and accurate astrometry and component parameters
from HST images (e.g. Lehar et al.~\cite{Lehar00}), it 
is now relatively easy to produce light curves. 
Arguably the best light curve available for quasar microlensing was
produced by the Optical Gravitational Lens Experiment (OGLE) using the same 
observing procedures as for their primary
Galactic microlensing experiment, combined with difference imaging to
analyze the results (Wozniak et al.~\cite{Wozniak00a}, \cite{Wozniak00b}).

The third problem is that we can never observe the ``unlensed'' source to get
a baseline from which to determine absolute magnifications.  This problem is
no worse than the blending problem for Galactic microlenses, in which the
flux of the lensed star is contaminated by flux from a nearby star 
(Di Stefano \& Esin~\cite{DiStefano95}) or many unresolved stars (pixel 
lensing, Crotts~\cite{Crotts92}).  It is
certainly true that there is no means of determining the absolute magnifications 
of the individual images because this is degenerate with the unknown flux of the quasar.  
However, by taking advantage of the spatial structure of quasars, it is 
possible to determine the true magnification ratios between the images in the
absence of microlensing.  The emission line, mid-infrared and radio 
emitting regions of quasars should all be large enough to average out
the effects of microlensing to allow the determination of the ``intrinsic''
flux ratios (e.g. Wyithe et al.~\cite{Wyithe02a} for Q2237+0305).  

The fourth problem is that the quasar lenses have sources that are time 
variable, making it necessary to separate intrinsic and microlensing 
variability. If the source is time variable and contaminating the
microlensing flux variations, then the light curves can be used to
determine the time delay between the images and the effects of the
intrinsic variability are eliminated by comparing the light curves
shifted by the time delay.  Moreover, the time delay measurement 
provides a direct estimate of the total surface density near the
lensed images (under the assumption that $H_0$ is known, see
Kochanek~\cite{Kochanek02}), which can be compared to the estimates
of the total or stellar surface density derived from analyzing 
the variability created by microlensing.  If the source is not 
variable, or the time delay is short compared to the microlensing time
scales, then it is unimportant for understanding the microlensing.

The fifth, and most significant problem, is the difficulty in interpreting
the quasar microlensing light curves.  Even the complex light curves produced by 
binary lenses (e.g. Mao \& Paczynski~\cite{Mao91}), 
are far simpler than those produced by the collective
effects of many stars.  The first observational studies of quasar microlensing
used semi-quantitative analyses
of the temporal widths of light curve peaks to estimate the size of the
accretion disk in the source quasar of Q2237+3035 (e.g.  Webster et al.~\cite{Webster91},
Wambsganss, Paczynski \&
Schneider~\cite{Wambsganss90}, Rauch \& Blandford~\cite{Rauch91}).  More
recent studies of the source structure focused on detailed analyses of 
``high magnification events,'' where the magnification pattern should have the generic 
asymptotic properties of a fold 
or cusp caustic (e.g. Yonehara~\cite{Yonehara01}, Shalyapin et al.~\cite{Shalyapin02}).  
General analyses of light curves have focused on estimates of their statistical properties.
In particular, 
Seitz \& Schneider~(\cite{Seitz94a}), Seitz, Wambsganss \& Schneider~(\cite{Seitz94b}) 
and Lewis \& Irwin~(\cite{Lewis96}) 
considered the auto-correlation functions of light
curves, Wyithe, Webster \& Turner~(\cite{Wyithe99})
considered the distributions of light curve
derivatives, and Lewis \& Irwin~(\cite{Lewis95}) considered the probability distributions
of the magnifications.   In all cases, the application of these statistical methods has been
to the four-image lens Q2237+0305 (Huchra et al.~\cite{Huchra85}) in order to
estimate the average microlens mass (e.g. Refsdal \& Stabell~\cite{Refsdal93},
Seitz et al.~\cite{Seitz94b}, Lewis \& Irwin~\cite{Lewis96}, 
Wyithe et al.~\cite{Wyithe00c}), 
the transverse velocity (Wyithe \& Turner~\cite{Wyithe01}), and the
source size and structure (e.g. Witt \& Mao~\cite{Witt94a},
Wyithe et al.~\cite{Wyithe00e}, Wyithe, Agol \& Fluke~\cite{Wyithe02a}).  
While these are reasonable statistical
estimators, they are difficult to apply to irregularly sampled, sparse
data and they lose information compared to the raw light curves because 
the statistics of the light curves are highly non-Gaussian.  The biggest
problem in using quasar microlensing for astrophysics remains the problem
of interpreting the data.

While many of the astrophysical applications of Galactic and quasar microlensing
analyses are similar, there is a fundamental difference in using the two
methods to study the dark matter problem.  In quasar microlensing, the behavior
of the light curves depends on both the density of the stars and the density
of the smoothly distributed matter. Moreover, the effects of the two density
components can be distinguished (e.g. Schechter \& Wambsganss~\cite{Schechter02}).
Simple studies of the dependence
of image flux ratios on image parities already suggest that in most quasar lenses
the stars must represent only a modest fraction of the total density 
(see Schechter \& Wambsganss~\cite{Schechter02}, Kochanek \& Dalal~\cite{Kochanek03}). 
This is very different from Galactic microlensing experiments which, even with infinite
resources, can only determine the density of the halo in compact
objects (stars, planets etc.).  The inference that the rest of the 
halo must be composed of smoothly distributed (particle) dark matter 
comes only from comparing the measured density to that inferred from
dynamical studies of the Galaxy.  With quasar microlensing, no additional
step is required.  The greater ability of quasar microlensing to address the
dark matter problem makes solving the problem of interpreting the data an
important one.   

In this paper we develop and demonstrate a method for obtaining physical information
from quasar microlensing data of arbitrary complexity and apply it to Q2237+0305.
We will simultaneously estimate the source 
velocities, source size, source structure, stellar mass function and stellar
surface density fraction needed to obtain statistically acceptable models of the
Q2237+0305 light curves measured by OGLE
(Wozniak et al.~\cite{Wozniak00a}, \cite{Wozniak00b}).  In doing so,
we also obtain model light curves that are consistent with the
observations.  We outline our approach in \S\ref{sec:method}, with
additional details on our method of computing microlensing magnification patterns
given in Appendix~\ref{sec:rayshoot}.  
Since the distribution of stars needed to reproduce the
available data is not unique, we introduce a Bayesian statistical 
method to estimate any physical variables of interest.  In \S\ref{sec:applic} we analyze the
OGLE light curves for Q2237+0305 to estimate the source velocity and average
stellar mass (\S\ref{sec:ve}), the source structure (\S\ref{sec:rshat}), the 
physical properties of the accretion disk and the mass of the black hole (\S\ref{sec:rs}),
the surface density of the stars (\S\ref{sec:kappa}), and the flux ratios of the images 
(\S\ref{sec:fluxrat}). In \S\ref{sec:lcurves} we survey some of the best fits to the 
light curves. Finally, in \S\ref{sec:discuss} we summarize the 
results and outline the potential future of quasar microlensing.

\section{A New Approach to Analyzing Quasar Microlensing Data \label{sec:method}}

Just as in the analysis of Galactic microlensing light curves (see Afonso et al.~\cite{Afonso00}
for a spectacular example), we will analyze
quasar microlensing light curves by finding configurations of stars and source
trajectories that reproduce the observations.  Because the stellar configurations
are complex, we search for good fits to the data by producing large numbers of 
random realizations of the light curves.  Then, using a Bayesian analysis of the 
the goodness of fit statistics for these model light curves, we estimate the 
values and uncertainties for any physical variable of interest.

We generate source plane magnification patterns using the ray-shooting method
(e.g. Schneider et al.~\cite{Schneider92}).  The technical details of our method,
which has a number of non-standard features,
are summarized in Appendix~\ref{sec:rayshoot}.  For our study of Q2237+0305 
we used fixed values from lens models for the mean convergence
$\kappa$ and shear $\gamma$ at the location of each image,
but considered models with a range 
of stellar mass fractions $\kappa_*/\kappa=1$, $1/2$, $1/4$ and $1/8$ 
where $\kappa_* \leq \kappa$ is the
surface density of the stars.  The stars are distributed randomly in position and are drawn
from a power-law mass function $dp/dM \propto M^{-x}$ over a finite mass
range $M_1 < M < M_2$.  We normalize our length scale by the Einstein radius
$\langle \theta_E \rangle$ corresponding to the average mass $\langle M \rangle$,
and parameterize the mass function by the exponent $x$ and the ratio between
the upper and lower masses $r = M_2/M_1$.  In the present calculation we use
either a Salpeter mass function ($x=2.35$) with a mass ratio $r=100$, 
or a ``mono-mass'' mass function in which all stars have the same mass ($r=1$).
Our standard magnification pattern was a square region spanning 
$40\langle \theta_E \rangle$ stored in a $2048^2$ array with a pixel
scale of $0.02\langle \theta_E \rangle$.
These scales were chosen so that we could make large numbers of statistically
independent trial light curves from a single magnification map.

In Galactic microlensing, stellar angular diameters are much smaller than the lens
Einstein radius, so the effects of finite source size are seen only during caustic
crossings (e.g.  Witt \& Mao~\cite{Witt94b}).  
For quasar microlensing there is less separation of the two scales,
making finite source sizes more important (e.g. Kayser, Refsdal \& Stabell~\cite{Kayser86},
Schneider \& Weiss~\cite{Schneider87}).  For a given source model, we
convolve the raw magnification pattern with the surface brightness model of the
source before computing the light curves.  The physical effects of the source size
are controlled by the ratio between the source size and the average Einstein radius,
$r_s/\langle\theta_E\rangle\propto r_s/\avgm^{1/2}$, so we assumed circular sources 
scaled by the average mass of the stars.  For length scale $r_s=\hat{r}_s\avgmhat^{1/2}$, 
we computed light curves for scales from $\hat{r}_s=10^{15}h^{-1}$~cm (slightly below
our pixel scale) to $10^{18}h^{-1}$~cm (somewhat above the average Einstein radius) 
in steps of $\Delta \log\hat{r}_s=0.25$.  We used either a Gaussian or a 
thin disk model for the surface brightness profile $I(R)$. 
The Gaussian model for the surface brightness,
\begin{equation}
       I(R) \propto \exp\left(-R^2/2 r_s^2\right)
\end{equation}
is the model usually used in microlensing studies.  For a comparison, we used
a standard model for an optically thick, pressure supported, absorption
opacity-dominated, thin accretion disk in which energy is released locally with a
black body spectrum (e.g. Shapiro \& Teukolsky~\cite{Shapiro83}).  
For a black hole of mass $M_{BH}$ and accretion rate
$\dot{M}$, the energy dissipation rate per unit area of the disk,
$3 G M_{BH} \dot{M}/8\pi R^3$, must equal the radiation losses of $\sigma T_s^4$,
so the disk surface temperature $T_s \propto R^{-3/4}$.  We will not include the correction
factor of $1-(3R_{BH}/R)^{1/2}$ to the dissipation rate near the last stable 
orbit of the black hole so as to avoid additional parameters.  For reasonably
narrow filters ($\Delta\lambda/\lambda \sim 15\%$ for the V-band) the surface
brightness of the disk   
\begin{equation}
   I(R) \propto \left[ \exp\left( (R/r_s)^{3/4} \right) -1 \right]^{-1}.
\end{equation}
simply tracks the black body spectrum.  The scale length $r_s$ is the
radius at which the disk surface temperature matches the effective wavelength
of the filter -- for V-band observations of Q2237+0305 (2000{\AA} in the rest frame), 
the  temperature at radius $r_s$ is $T_s(r_s) \simeq 70000$~K.  The thin disk
model can be used to make self-consistent predictions for the wavelength dependence 
of the microlensing effects because the radius scales with photon wavelength
as $r_s \propto \lambda^{4/3}$. 

The light curves produced by the two models weight the magnification pattern
very differently.  On small scales, $R<r_s$, the Gaussian model has nearly constant 
surface brightness while the black body model is a centrally peaked power
law, $I(R) \sim R^{-3/4}$.  On large scales, $ R > r_s$, the Gaussian model
cuts off much more sharply than the black body model.  We will
consider only circular (face on) disks to avoid introducing two additional
parameters for the inclination and orientation of the disk.  This means that
estimates of the scale length will tend to be underestimates. Crudely, 
microlensing measures the area of the source rather than the radius, 
so for circular scale length $r_{s,circ}$ the true scale length $r_{s,true}$ of a
disk with axis ratio $q$ is roughly $r_{s,true} \simeq r_{s,circ}/(1-q)^{1/2}$. 
 
Once we have the convolved magnification pattern, we can choose an initial point
${\bf u}_0$ and an effective velocity 
${\bf v}_e= v_e\left(\cos\Theta,\sin\Theta \right)$ for the trajectory to
compute the magnification as a function of time.  
We make two simplifications in generating the light curves. First, we neglect
the internal motions of the stars in the lens galaxy and use fixed
magnification patterns. Studies of the effects of moving stars 
(e.g. Kundic \& Wambsganss~\cite{Kundic93},
Schramm et al.~\cite{Schramm93}, Wyithe, Webster \& Turner~\cite{Wyithe00a})
generally found that their effects were difficult to statistically distinguish
from a simple, static magnification pattern.
Secondly, we regard the
trajectory directions ($\Theta$) as independent, uniformly distributed random
variables for each image.  We experimented with the effects using the
same $\Theta$ for all images and found that it had little effect on the
results. Moreover, the neglected
internal motions of the stars ``randomize'' the trajectories,
making perfectly locked trajectories unphysical without the inclusion
of the stellar motions. 
For each trajectory
we compute the change in 
magnitudes, $\delta\mu^\alpha(t)$, produced by microlensing image $\alpha$, 
relative to the mean magnification for the image. 

\subsection{Fitting the Data \label{sec:chidef} }

The data consists of a series of magnitude measurements $m^\alpha_i$ for image $\alpha$ at 
epoch $i$ with uncertainties $\sigma_{\alpha,i}$.  These magnitudes are a combination of 
the source magnitude at that epoch $S_i$, the local mean magnification for the image 
$\mu^\alpha$ (as a magnitude), any offsets in the magnitude due to extinction, substructure 
or other systematic effects on the image fluxes $\Delta\mu^\alpha$, and the time varying 
change in the magnification due to microlensing $\delta\mu_i^\alpha$ relative to the local mean,
\begin{equation}
     m^\alpha_i = S_i + \mu^\alpha + \Delta\mu^\alpha + \delta\mu_i^\alpha
                = S_i + \mu^\alpha_{tot,i}.
\end{equation}
We measure the goodness of fit with a $\chi^2$ statistic, 
\begin{equation}
    \chi^2 = \sum_\alpha \sum_i 
  \left[ { m^\alpha_i-S_i -  \mu^\alpha_{tot,i} 
         \over \sigma_{\alpha,i} } \right]^2.
\end{equation}
In addition to the microlensing magnification curves, $\delta\mu_i^\alpha$, for each image, 
the model parameters are the source flux $S_i$, and the offsets $\Delta\mu^\alpha$ from
the mean magnification.  If there is a significant time delay between the images, then
we would need to include the appropriate temporal offsets between the light curves.

The source magnitude must be determined for each individual model since it is not a
direct observable.  We can do so either by estimating it from the data for each 
epoch or by assuming a parameterized model for its variation with time.  If we 
estimate it from the data for each epoch, which we will call a ``non-parametric''
model, we solve $\partial\chi^2/\partial S_i=0$ to find that
\begin{equation}
     S_i = \left[ \sum_\alpha { m^\alpha_i-\mu^\alpha_{tot,i} \over \sigma_{\alpha,i}^2 }\right]
           \left[ \sum_\alpha { 1  \over \sigma_{\alpha,i}^2 }\right]^{-1}.
     \label{eqn:chiall1}
\end{equation} 
The $\chi^2$ statistic then reduces to a sum over the $N(N-1)/2$ possible difference light 
curves of the $N$ images,
\begin{equation}
    \chi^2 = \sum_\alpha \sum_{\beta < \alpha} \sum_i 
          \left[  { \left( m^\alpha_i-\mu^\alpha_{tot,i} \right) -
                    \left( m^\beta_i-\mu^\beta_{tot,i} \right)
         \over \sigma_{\alpha\beta,i} } \right]^2. 
     \label{eqn:chiall2}
\end{equation}
The errors $1/\sigma_{\alpha\beta,i}^2$ are the product of the $N-2$ errors excluding 
images $\alpha$ and $\beta$ divided by the sum of all the
exclusive permutations of $N-1$ errors.  For example, if we have 4 images labeled A-D, the 
weighting for the A/B difference light curve is
\begin{equation}
        { 1 \over \sigma_{AB,i}^2 } = { \sigma_{C,i}^2 \sigma_{D,i}^2 \over
            \sigma_{A,i}^2\sigma_{B,i}^2\sigma_{C,i}^2+
            \sigma_{A,i}^2\sigma_{B,i}^2\sigma_{D,i}^2+
            \sigma_{A,i}^2\sigma_{C,i}^2\sigma_{D,i}^2+
            \sigma_{B,i}^2\sigma_{C,i}^2\sigma_{D,i}^2 }.
\end{equation}
While statistically optimal, the actual source behavior can be unphysical if we are
confident that the intrinsic variability and microlensing effects have different time
scales.  For example, suppose image A is crossing a caustic and has a peak, while 
image B has more or less constant flux.  If we have a poor model for the microlensing 
light curves with a peak at neither A nor B, then the source will be given a peak which 
is half the amplitude of the observed peak.  If we are confident that the source should
be varying slowly, then the {\it a priori} probability of the source conspiring to
mimic part of the microlensing peak is low. 
We can force the source to show little correlation with shorter time-scale microlensing
variability by using a parametric model for the source.  For example, a source described
by a polynomial $S_i = p_0 + p_1 t_i + p_2 t_i^2 \cdots$ function of the epoch $t_i$ 
leads to simple linear equations $\partial \chi^2/\partial p_i=0$ for the source parameters.
Parameterized source models also allow us to fit the light curves of one image at a time.
In particular, if we assume that the source has a nearly constant magnitude $S_0$
with random magnitude fluctuations of $\sigma_0$, then we can fit the light curve of a
single image as
\begin{equation}
      \chi^2_\alpha = \sum_i { \left( m_i^\alpha - S_0 - \delta\mu_i^\alpha \right)^2
             \over \sigma_{\alpha,i}^2 + \sigma_0^2 }.
    \label{eqn:chi1}
\end{equation}
Analyzing a single image allows for far more rapid calculations than joint analyses
of four images because it avoids the combinatoric explosion we discuss in
\S2.2.  We will call these ``parametric'' models.

Although there is no theoretical problem with  including
measurements (e.g. extinction estimates) or constraints (e.g. the relative macro
magnifications must be correct to some accuracy) on the magnitude offsets, 
we decided that for our present study we would use only the time variability of the
images to constrain the models.  This means that we solve for the optimal value of 
the offsets, $\Delta\mu^\alpha$, for every trial light curve.  
If our time series is sufficiently long, so that it averages over many Einstein  
radii of the microlensing pattern, then these estimates of the offsets from fitting
the light curves should converge to their true value.  Otherwise, they will show
significant scatter depending on whether the light curve lies in a region of higher
or lower than average microlensing magnification.

\subsection{Dealing With The Combinatoric Explosion \label{sec:combine}}

The probability of a randomly drawn microlensing magnification
curve leading to a reasonable fit to the OGLE light curves is small and we cannot try every
possible trajectory for a broad range of physical parameters.  For this study we used  
magnification patterns with an outer scale of $40\langle\theta_E\rangle$ and
dimensions of $2048 \times 2048$ pixels, leading to an inner, pixel scale 
of $0.02\langle\theta_E\rangle$.  For a compact source and
a light curve with a caustic crossing feature, testing all possible trial light curves
for a single pattern, source size and effective velocity would require of order 
$10^{14}$ trials.\footnote{  We note, however, that there is a trick using Fourier transforms 
  to efficiently check all possible starting points 
  even for very large numbers of data $N_{dat}$.  For a fixed source velocity
  and angle, the data points imply a spatial filter consisting of delta 
  functions $\delta({\bf u} -{\bf u}_i)$ located at spatial positions from
  the first point that are determined by the effective source velocity ${\bf v}_{e,i}$
  and the elapsed time, ${\bf u}_i = {\bf v}_{e,i} \Delta t_i$.  The $\chi^2$ for all
  possible ray starting points is then formed from the convolution of this ``beam'' 
  with the magnification pattern and its square.  For magnification patterns with
  $N_{pix}$ pixels, this approach requires of order $O(N_{pix} \ln N_{pix})$ operations
  rather than the order $O(N_{pix} N_{dat})$ operations that a direct search would need.    
  Unfortunately, the convolutions must be repeated for each trial velocity.  For very
  large data sets, this technique could be used to prefilter the magnification patterns
  at low resolution to locate regions deserving higher resolution searches. }
If we want study more than one image over a broad range of effective
velocities, source sizes and physically different magnification patterns, then we
are forced to use Monte Carlo methods to search a random sampling of the trajectories.
In practice we find that for fitting a single image of Q2237+0305 assuming a constant source 
with random intrinsic fluctuations of $\sigma_0=0.05$~mag that approximately one in every 
$N \simeq 100$ trial realizations will produce a fit with $\chi^2/N_{dof} \ltorder 3$
where $N_{dof}$ is the number of degrees of freedom.  
Obviously a much smaller fraction produce fits with $\chi^2/N_{dof} \simeq 1$.

The problem explodes when we try to fit more than one light curve simultaneously.
Crudely, if we fit 2, 3 or 4 light curves simultaneously we would
expect that it would take $N^2 \simeq 10^{4}$, $N^3\simeq 10^{6}$ or
$N^4 \simeq 10^{8}$ trials to produce equally good fits to all the images simultaneously.  
At least when using the non-parametric method, the scaling is less extreme because
there are so many degrees of freedom in the source.  In practice, finding a fit for two
images using the non-parametric method is not much harder than finding a fit for one 
image with the simpler parametric method.  It is possible to find reasonable
four-image solutions in $N\simeq 10^6$ trials.  
We speed the process of finding good realizations in two ways.  

First, we set a 
threshold, $\chi_{max}^2$, on the value of the $\chi^2$ statistic, and assume that any 
light curve exceeding this value (and any local perturbations to it) should have
zero statistical weight in our analysis.  We then note that as we add data points to the 
determination of a $\chi^2$ statistic, the statistic can only increase in absolute value.  
We take advantage of this by computing the $\chi^2$ using the data points in a random temporal
order and stopping the calculation as soon as $\chi^2 > \chi^2_{max}$.  If we set 
$\chi^2_{max} = 3 N_{dof}$ (for the parametric models, $5N_{dof}$
for the non-parametric models), then the vast majority of trials light curves are
disposed of based on a small fraction of the data points.  Because nearby points of both
the light curves and the magnification patterns tend to be similar, while well-separated
points tend to be dissimilar, random ordering of the data allows much faster rejection 
of a trial light curve than sequential ordering.

Second, for trial light curves which have $\chi^2<\chi^2_{max}$, we locally optimize the
parameters (the starting points ${\bf u}_0$ and the directions at fixed effective velocity $\Theta$) of the 
curves to minimize the $\chi^2$.  This step helps considerably in finding good solutions
given our inability to try every possible set of initial conditions in our magnification
pattern.  We get a fair random sampling of the global initial conditions, but allow for
a local optimization since we cannot perform the fine sampling needed to try every initial
condition.  The optimization step means that we need to keep our threshold $\chi^2_{max}$
sufficiently high so that typical optimizations of cases above the threshold would not 
reduce the $\chi^2$ to the point where the trials become statistically significant.
 
There is some risk that these modifications can create biases in the results.  For example,
in regions with complex caustic structures the source trajectory requires better alignment
with the magnification pattern in order to fit the data than in regions with less complex
structures.  Hence, the combination of an initial threshold followed by local optimization
could bias our results against finding solutions in the complex regions.  While it was
not computationally feasible to conduct our complete model survey without a threshold,
we did test specific cases and found no evidence for the procedures introducing a
bias.

\subsection{Parameter Estimation \label{sec:bayes} }

We use Bayesian methods for parameter estimation based on comparing large
numbers of trial light curves to the observed data.  The statistical
properties of the light curves expected for each image depend on the
local magnification tensor ($\kappa$ and $\gamma$), the local properties
of the stars ($\kappa_*$, $\langle M \rangle$, $x$ and $r$), the structure of the 
source (Gaussian or thin disk, $r_s$) and the effective velocity
of the source ${\bf v}_e= v_e\left(\cos\Theta,\sin\Theta \right)$.  
We will collectively refer to these physical
parameters as ${\bf \xi_p}$.  For any given set of physical parameters
we generate large numbers of source trajectories described by their
starting points (${\bf u}_0$) and directions ($\Theta$).  We regard
these trajectory parameters, which we will collectively refer to as
${\bf \xi_t}$, as nuisance parameters that we will project out of the
likelihoods.

For each trial light curve we obtain a goodness of fit defined by the $\chi^2$
statistics introduced in \S2.1.  Our next step is to define the relative 
likelihoods of the light curves given the $\chi^2$ values.  Using a standard 
maximum likelihood estimator, like $P(D|{\bf \xi_p},{\bf \xi_t})=\exp(-\chi^2/2)$, 
works poorly because we are comparing the probabilities of completely
different light curves rather than models related to each other by continuous
changes of parameters.  We would expect even ``perfect'' model light curves
to have $\langle\chi^2\rangle \simeq N_{dof} \pm (2N_{dof})^{1/2}$, so only $\chi^2$ 
differences of order $(2N_{dof})^{1/2}$ indicate whether one
light curve is superior to another.  For this reason we base
our likelihoods on the probability of obtaining a given value of $\chi^2$
for data with  $N_{dof}$ degrees of freedom,
\begin{equation}
   P(\chi^2|N_{dof})={dP \over d\chi^2 } \propto \chi^{N_{dof}-2} \exp\left(-\chi^2/2 \right).
   \label{eqn:chidist}
\end{equation}
The second problem is that we are fitting data with a large number of degrees of freedom 
($N_{dof}=290$ for the simultaneous fits to all four images of Q2237+0305 discussed in \S3),
so our $\chi^2$ estimates are very sensitive to small errors in 
the magnitude uncertainties of the light curves.  It takes only a 4\% shift in
the magnitude uncertainties to produce a $(2N_{dof})^{1/2}$ change in $\chi^2$
when $N_{dof}=290$. 

We control this problem by allowing for uncertainties in the magnitude errors 
$\sigma_\alpha,i$.  If we scale the magnitude errors by the factor $f$, 
then the value of $\chi^2$
changes to $\chi_f^2=\chi^2/f^2$ with distribution 
$P(\chi_f^2|N_{dof})=P(\chi^2/f^2|N_{dof})/f^2$.  By averaging over $f$,
weighted by some prior $P(f)$ for our level of uncertainty in the errors,
we can obtain estimates for the relative probabilities of the models that
are insensitive to errors in the magnitude uncertainties.  We set the
magnitude errors to be the quadrature sum of the OGLE uncertainties and
$0.05$ mag, and found that our best fit models had $\chi^2 \simeq 200$
for $N_{dof}=290$.  This suggests that we overestimated the magnitude
errors by at least 20\%, and that we can assume  $0 \leq f \leq 1$.
Since the data contains real measurement errors, $P(f)$ must approach
zero as $f\rightarrow 0$.  For simplicity we adopt
$P(f) \propto f$ for $0 \leq f \leq f_0=1$, in 
which case the weighted average of $P(\chi^2_f|N_{dof})$ over $f$ becomes  
\begin{equation}
   P(\chi^2) \propto \Gamma\left[ { N_{dof}-2 \over 2 }, { \chi^2 \over 2f_0^2 } \right],
   \label{eqn:probfunc}
\end{equation}
where $\Gamma[a,b]$ is an incomplete Gamma function.\footnote{We experimented with 
  other plausible choices and found they had no significant effects on our results. 
  For example, using a range from $f_1 \leq f \leq f_0$ gives the difference of
  two Gamma functions, 
  $\Gamma[(N_{dof}-2)/2,\chi^2/2f_0^2] -\Gamma[(N_{dof}-2)/2,\chi^2/2f_1^2]$.
  This function gives a $\chi^2/f_1^2$ distribution for $\chi^2/f_1^2<N_{dof}$, a
  $\chi^2/f_0^2$ distribution for $\chi^2/f_0^2>N_{dof}$, and a plateau
  in the intermediate region where distinguishing models depends more on
  the uncertainty in the errors used to construct the $\chi^2$ statistics
  than on the any differences between the light curve realizations. }
This expression has the ``correct'' properties for estimating the relative
probabilities of light curve realizations.  First, like the $\chi^2$ distribution, light 
curve realizations must have $\chi^2$ differences comparable to $(2N_{dof})^{1/2}$ 
before they have significantly different relative probabilities.  Second,
when $\chi^2$ is larger than $N_{dof}$, it simply becomes a $\chi^2$ distribution
set by the maximum plausible error $f_0$ and with an unimportant reduction  
in the number of degrees of freedom.  Third, when the $\chi^2$ is smaller than
$N_{dof}$, the likelihood of the models rises, with $P(0)/P(N_{dof})\simeq 2$,
rather than falling as it does for the true $\chi^2$ distribution (Eqn.~\ref{eqn:chidist}).  
When we find
models with $\chi^2 < N_{dof}$, it is probably because we have overestimated
the magnitude errors rather than because we have over fit the data.  In summary,
the advantage of this likelihood estimator is that it benignly handles the
problem of systematic uncertainties in the $\chi^2$ estimators even when
$N_{dof}$ is large.  Our approach is conservative because it will overestimate
the uncertainties in any results provided the true errors correspond to
the region with $f \leq f_0=1$. 

Using Bayes' theorem, the probability of the parameters given the data is  
\begin{equation}
     P({\bf \xi_p},{\bf \xi_t}|D) \propto 
     P( D | {\bf \xi_p},{\bf \xi_t}) P({\bf \xi_p})P({\bf \xi_t})
\end{equation}
where $P({\bf \xi_p})$ and $P({\bf \xi_t})$ describe the prior probability 
estimates for the physical and trajectory variables respectively, and
$P(D|{\bf \xi_p},{\bf \xi_t})=P(\chi^2)$ as defined in Eqn.~\ref{eqn:probfunc}.
All Bayesian parameter estimates are normalized by the
requirement that the total probability is unity, 
$\int d{\bf \xi_p}d{\bf \xi_t} P({\bf \xi_p},{\bf \xi_t}| D)=1$.  
We assume that the trajectory starting
points and directions are uniformly distributed and that they are nuisance
variables.  We obtain the probability distributions for the more interesting
statistical parameters by marginalizing over the trajectory variables
\begin{equation}
     P({\bf \xi_p}|D) \propto \int P({\bf \xi_p},{\bf \xi_t}|D) d{\bf \xi_t}.
\end{equation}
In practice we sum the probabilities for our random sampling of trajectories,
which is equivalent to using Monte Carlo integration methods to compute the
integral over the space of all possible trajectories.  The sum over the
random trajectories will converge to the true integral provided we make
enough trials.

For our present study we assumed that the values of $\kappa$ and $\gamma$
are known exactly from lens models.  We studied a range of values for the
fraction of the surface density composed of stars with a logarithmic
prior $P(\kappa_*) \propto 1/\kappa_*$.  We considered discrete trials
of the different mass function parameters ($x$ and $r$) and the two
source structures with all the cases given equal prior likelihoods.
We used a logarithmic prior $P(\hat{r}_s) \propto 1/\hat{r}_s$
for the scaled source size where $r_s = \hat{r}_s \avgmhat^{1/2}$.
We also use the source velocity scaled by the average mass of the
lenses $\hat{v}_e$, where $v_e =\hat{v}_e \avgmhat^{1/2}$, as our
computational variable.  We used a logarithmic prior $P(\hat{v}_e) \propto 1/\hat{v}_e$ 
for the scaled source velocity, which corresponds to a logarithmic prior
for the average stellar mass $\avgm$ combined with any prior for the 
distribution of physical velocities.

Ultimately we would like to obtain an estimate of the average microlens mass,
$\avgm$, which can be
done by combining the likelihood function $P(\hat{v}_e|D)$ for $\hat{v}_e$ we obtain from
fitting the light curves with a prior probability estimate $P(v_e)$ for the
true effective source velocity $v_e$, such that
\begin{equation}
    P(\avgmhat|D) \propto \int dv_e P(v_e)
       P\left( \hat{v}_e= v_e\avgmhat^{-1/2}|D \right).
   \label{eqn:massest}
\end{equation}
The effective source velocity, $v_e$, defined to be the change in the (proper) source position
per unit of time measured by the observer, is a distance-weighted combination of the 
(physical) transverse velocities of the observer, ${\bf v}_o$, lens, ${\bf v}_l$, and 
source, ${\bf v}_s$, respectively, is
\begin{equation}
   {\bf v}_e = {{\bf v}_o \over 1+z_l} { D_{LS} \over D_{OL} } 
             - {{\bf v}_l \over 1+z_l} { D_{OS} \over D_{OL} } 
             + {{\bf v}_s \over 1+z_s}
\end{equation}
(e.g. Kayser, Refsdal \& Stabell~\cite{Kayser86}).  The transverse velocity of
the observer is simply the projection of the heliocentric CMB dipole velocity
${\bf v}_{CMB}$ onto the lens plane,
\begin{equation}
  {\bf v}_o={\bf v}_{CMB} - ({\bf v}_{CMB} \cdot \hat{\bf z}) \hat{\bf z} 
\end{equation}
where $\hat{\bf z}$ is a unit vector in the direction of the lens.  With an
amplitude of $v_{CMB}=387$~km/s (e.g. Kogut et al.~\cite{Kogut93}), the 
observer's motion will be important for some lenses, and unimportant for
others, depending on the location of the lens.  The motions of the lens 
and source galaxies are assumed to match that expected from theoretical
estimates of peculiar velocities.  We model the (one-dimensional) peculiar
velocity dispersion as 
$\sigma_{pec}/(1+z)^{1/2}f(\Omega_0,\Lambda_0,z)/f(\Omega_0,\Lambda_0,0)$
and we use the approximations for the growth factor $f$ from
Eisenstein \& Hu~(\cite{Eisenstein99}).  Nagamine, Hernquist \& Springel
(2003, private communication) find that $\sigma_{pec}\simeq 235$~km/s
for a standard concordance cosmology.   
The final contribution to the effective source motion is the velocity
dispersion of the stars in the lens galaxy, $\sigma_*$.  Because we 
use fixed magnification patterns, we cannot treat this component 
exactly. However, experiments by Wyithe, Webster \& Turner~(\cite{Wyithe00a}) found 
that for the statistics of light curve derivatives they could model the effects 
of the stellar velocity dispersion as a bulk velocity scaled by an efficiency 
factor $0.8 \ltorder \epsilon \ltorder 1.3$ that depended on the local values
of $\kappa$ and $\gamma$.  

In order to define the probability distribution of source effective velocities,
$P(v_e)$, we divide the various terms into Gaussian and fixed components.
We treat the unknown peculiar velocities of the lens and the source as
Gaussian distributed variables summing them in quadrature to give a total
one-dimensional source plane velocity dispersion of 
\begin{equation}
  \sigma_e^2= \left[ {\sigma_{pec}(z_l) \over 1+ z_l } { D_{OS} \over D_{OL} }\right]^2
    + \left[ {\sigma_{pec}(z_s) \over 1+ z_s }\right]^2.
   \label{eqn:vsig}
\end{equation}
We treat the fixed projection of the CMB velocity onto the source plane and
the stellar velocity dispersion as constant velocities, summing the two 
contributions in quadrature to give an average velocity of
\begin{equation}
  \bar{v}_e^2
   = \left[ {v_{CMB} \over 1+ z_l } { D_{LS} \over D_{OL} }\right]^2
    + 2 \left[ {\epsilon \sigma_* \over 1+ z_l } { D_{OS} \over D_{OL} }\right]^2.
   \label{eqn:vbar}
\end{equation}
We will assume $\epsilon=1$ as it has only modest effects on our estimates of
the average microlens mass $\langle M \rangle$.  We treat the stellar dispersion
as a fixed velocity component rather than as a Gaussian variable because it is
meant to model the collective, average effect arising from the random motions
of many stars.  
If we then average over the angle between the random Gaussian components and the
fixed component, the probability distribution for the magnitude of the effective
source plane velocity becomes
\begin{equation}
    P(v_e) = { v_e \over \sigma_e^2 } 
       \hbox{I}_0 \left[ { v_e \bar{v}_e  \over \sigma^2 } \right]
       \exp\left( - { v_e^2 +  \bar{v}_e^2 \over 2 \sigma^2 } \right)
\end{equation}  
where $I_0(x)$ is a modified Bessel function.
The root-mean-square (rms) source velocity,
$\langle v_e^2 \rangle^{1/2} = (\sigma_e^2+\bar{v}_e^2)^{1/2}$, is the same
as would be obtained treating all the velocities as Gaussian distributed
variables, but the Gaussian model would have broader wings.

\begin{figure}
\centerline{\psfig{figure=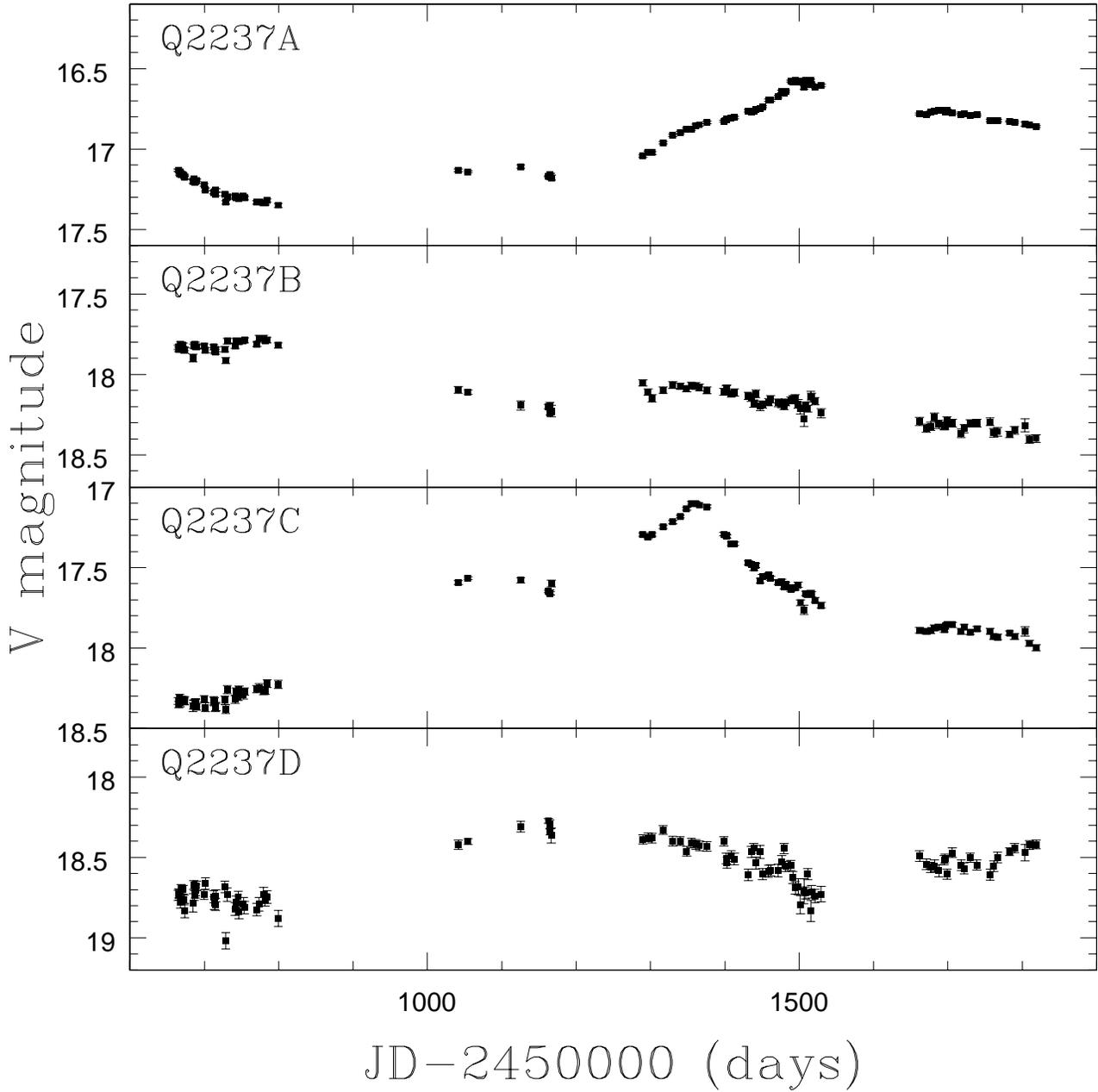,width=7.0in}}
\caption{ The OGLE V-band light curves for images A (top) to D (bottom) of Q2237+0305.
  Points separated by less than four hours have been combined.  The vertical scale
  of each panel is fixed to 1.5~mag.  The scatter between adjacent points suggests
  that the formal error bars shown here should be enlarged by $0.02$, $0.03$, $0.04$
  and $0.05$~mag for the A-D images respectively. 
  }
\label{fig:data2237}
\end{figure}

\begin{figure}
\centerline{\psfig{figure=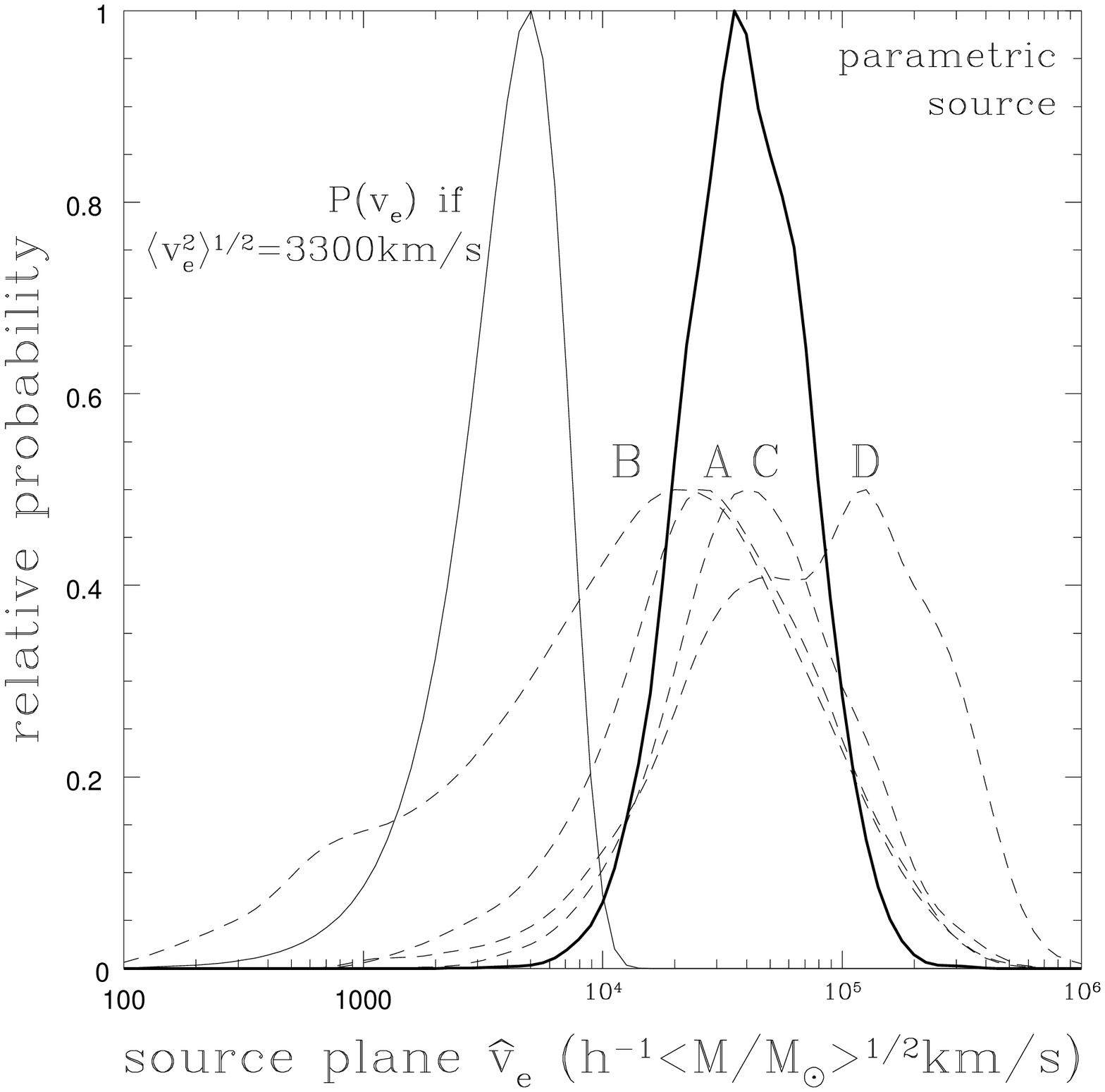,width=3.25in}
            \psfig{figure=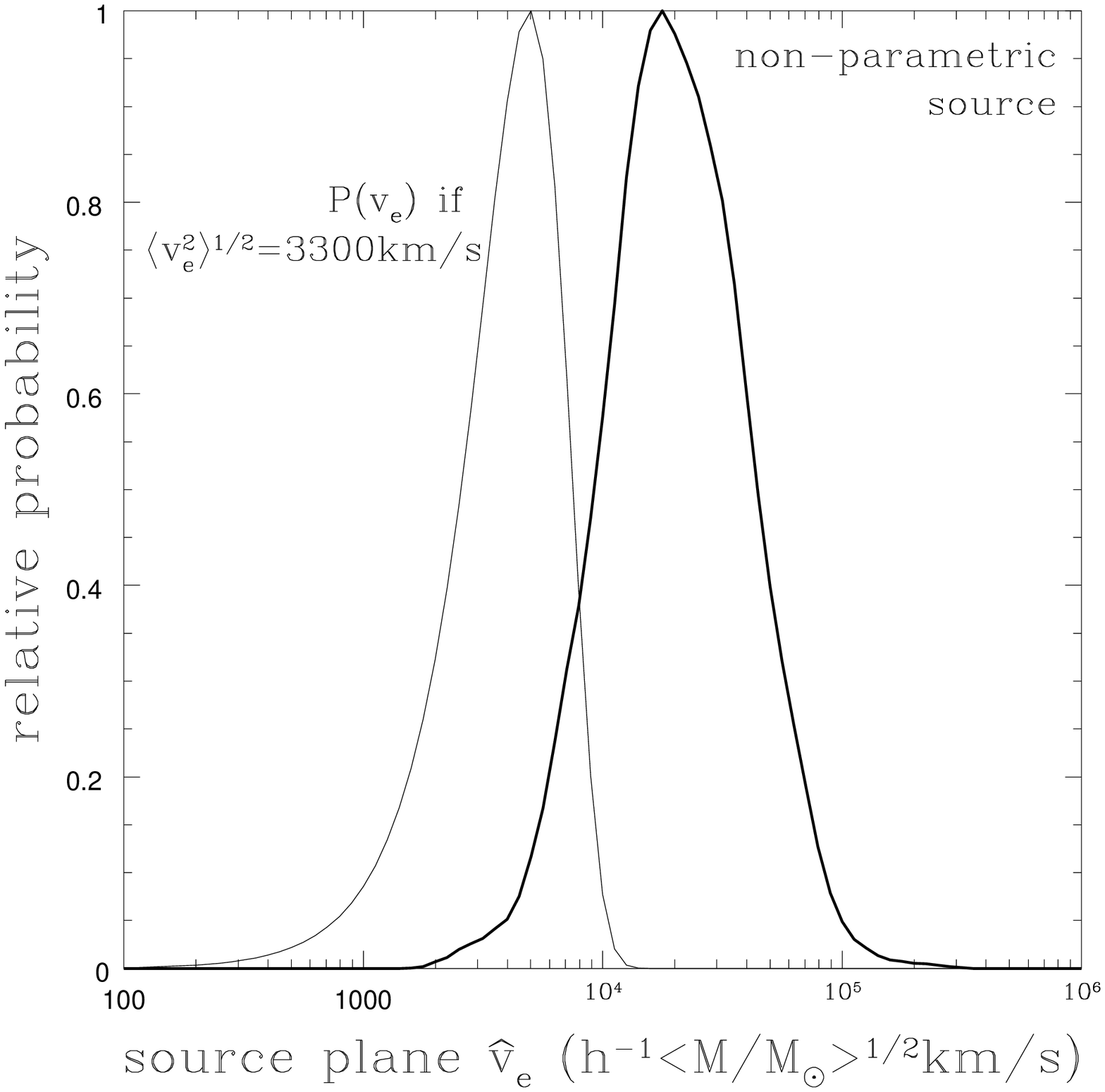,width=3.25in}}
\caption{ Probability distributions for the effective source plane velocity 
  $\hat{v}_e$ using the parametric (left) or non-parametric (right) source
  models.  The heavy solid curve normalized to a peak of unity shows the
  joint estimate from all four images. The light solid line shows our estimated 
  probability distribution $P(v_e)$ for the true source plane velocity $v_e$. 
  The offset of the two curves allows us to estimate the average 
  mass $\protect{\avgm}$.   For the parametric model, the dashed lines
  normalized to a peak probability of one-half show the independent estimates 
  from the A-D images. 
  }
\label{fig:veff}
\end{figure}

\section{Interpreting Q2237+0305 \label{sec:applic} }

We will use only the OGLE monitoring data for Q2237+0305 (Wozniak et al.~\cite{Wozniak00a},
\cite{Wozniak00b}) because
it covers a relatively long period (3 years) with relatively dense coverage (222 usable
points). Other data sets cover longer time periods with lower sampling rates (e.g.
Corrigan et al.~\cite{Corrigan91}, Ostensen et al.~\cite{Ostensen96}) or 
shorter periods with higher sampling rates (e.g. Alcalde et al.~\cite{Alcalde02}).  
We will not make use of the information on the true flux ratios in the absence
of microlensing derived from monitoring the CIII] emission line (Racine~\cite{Racine92},
Saust~\cite{Saust94}, Lewis et al.~\cite{Lewis98}), radio observations 
(Falco et al.~\cite{Falco96}) or
mid-infrared observations (Agol, Jones \& Blaes~\cite{Agol00}, Wyithe, Agol \&
Fluke~\cite{Wyithe02a}).  While adding this additional information poses no 
theoretical problems, we want to avoid any complications associated with
differences in filters, zero-points or extinction in this first analysis.  

On short 
time scales the light curve variations are smooth, so for faster calculation we 
averaged data spanning less than 4 hours into a single point, leaving 103 data points. 
Fig.~\ref{fig:data2237} shows the resulting light curves of the four images.  From the
scatter between adjacent points in the raw light curve, we estimated that our averaged 
light curves have larger
uncertainties than their formal errors.  Modeled as a term to be added in quadrature
with the formal errors, we found additional scatter of $0.02$, $0.03$, $0.04$, and
$0.05$~mag for the A, B, C and D images respectively.  To compensate for these and
any other systematic effects, we added $\sigma_0=0.05$~mag additional error in quadrature
to the uncertainties used to define the $\chi^2$ statistics.  As discussed in
\S2.3 (Eqn.~\ref{eqn:probfunc}), we then define the probabilities to allow for this being an
overestimate. 
We fixed the parameters of the macro model to those for a standard model consisting
of a singular isothermal ellipsoid (SIE) in an external shear field with no weight
assigned to reproducing the image flux ratios.
This gave ($\kappa$,$\gamma$) of ($0.394$, $0.395$), ($0.375$, $0.390$),
($0.743$, $0.733$) and ($0.635$, $0.623$) for images A, B, C and D
respectively.  These values are similar to those used in earlier studies (see
the summary in Wyithe et al.~\cite{Wyithe02b}).  

For an $\Omega_0=0.3$ flat cosmological model with $H_0=100h^{-1}$~km/s/Mpc, the angular 
diameter distances are $D_{OL}=113h^{-1}$~Mpc, $D_{OS}=1223h^{-1}$~Mpc and 
$D_{LS}=1180h^{-1}$~Mpc given the lens and source redshifts of $z_l=0.0394$ and 
$z_s=1.695$ (Huchra et al.~\cite{Huchra85}).  The source-plane Einstein radius 
of a star with the average mass, $\avgm$, is
\begin{equation}
   \langle \theta_E \rangle = 
     D_{OS} \left[ { 4 G \avgm \over c^2 D_{OL} } { D_{LS} \over D_{OS} } \right]^{1/2} 
     = \left( 1.54 \times 10^{17} \right)
       \left[ { \avgm \over M_\odot } \right]^{1/2} h^{-1} \hbox{cm},
\end{equation} 
and an effective source plane velocity of approximately 
$5\times 10^4 h^{-1} (\avgm/M_\odot)^{1/2}$~km/s is needed to cross the Einstein radius in
one year.   As first noted by Kayser \& Refsdal~(\cite{Kayser89}), the effective source 
velocity is dominated by the motion of the lens and its stars. 
The projection of the CMB dipole, ${\bf v}_{CMB} \simeq (-52,-23)$~km/s East
and North respectively, is quite small for Q2237+0305, so its contribution to the
effective source plane velocity of ${\bf v}_{e,CMB}=(-530,-230)$~km/s  can be
ignored despite the large boost from the distance ratios.  The peculiar velocity
of the source is unimportant because even if it were the same magnitude as that
of the lens galaxy, it does not get any boost from the distance ratios.  The
measured stellar velocity dispersion of the bulge is $\sigma_* = 215$~km/s 
(Foltz et al.~\cite{Foltz92}), roughly equal to the rms peculiar velocity of
the lens galaxy.  As a result, the mean velocity of 
$\bar{v}_e=2460$~km/s and the mean velocity dispersion of  $\sigma_e=2250$~km/s
are nearly identical and the total rms velocity is 
$\langle v_e^2\rangle^{1/2} = 3330$~km/s (see Eqns.~\ref{eqn:vbar} and \ref{eqn:vsig}).  
Changes in the efficiency factor
for the effects of the stellar velocity dispersion from $\epsilon=1$ produce 
small changes
in the estimated velocities.  The typical Einstein radius crossing time
is approximately $15 h^{-1} \avgmhat^{1/2}$~years.

\begin{figure}
\centerline{\psfig{figure=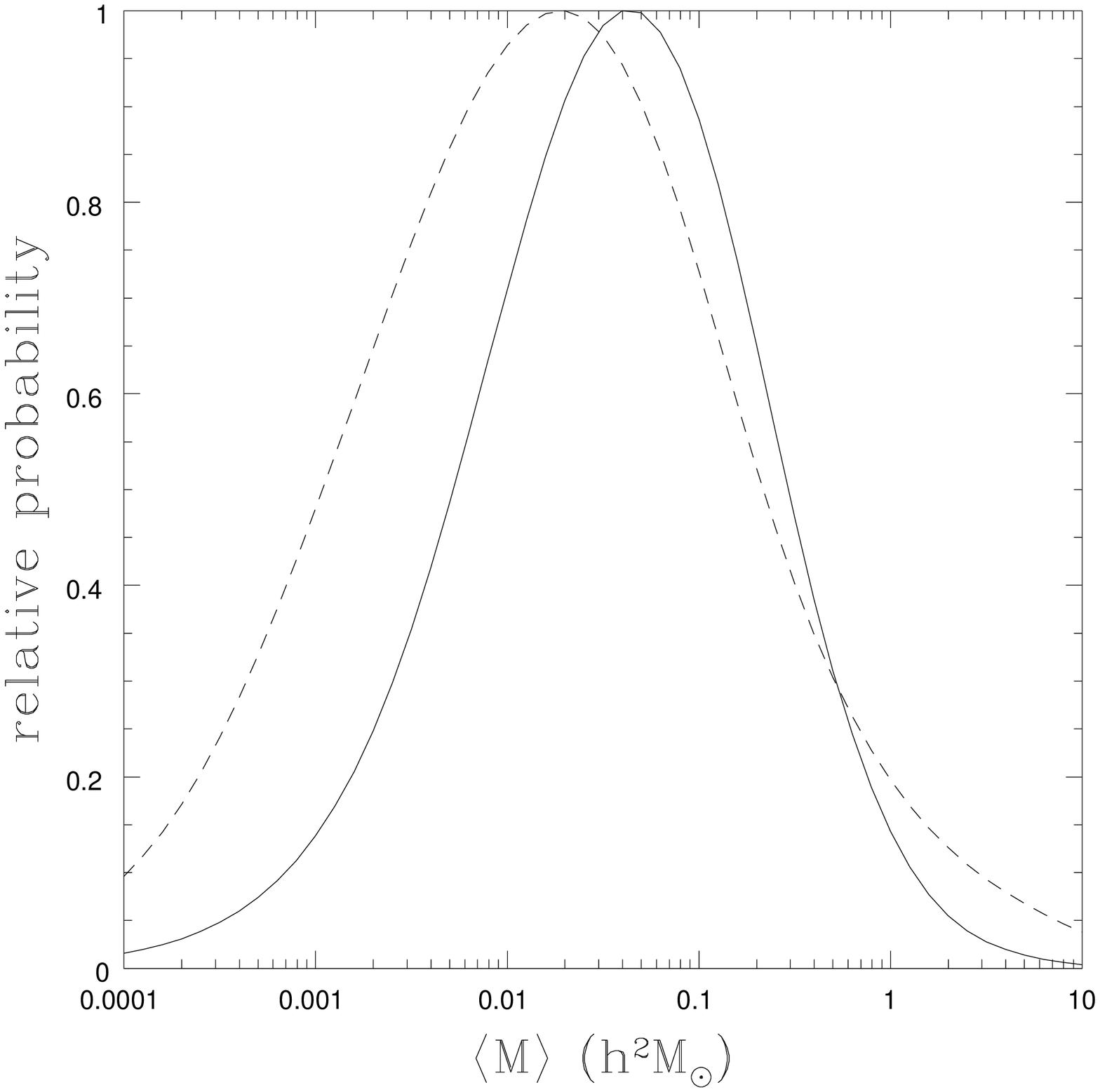,width=3.20in}}
\caption{ The probability distributions for the average mass $\protect{\avgm}$
  using the parametric (dashed curve) and non-parametric (solid curve) source 
  models.   The uncertainties are broad because $\protect{\avgm}\propto v_e^{-2}$.
  The shift in the mass scale between the parametric and non-parametric results
  is a consequence of the shift between their effective velocity distributions
  in Fig.~\protect{\ref{fig:veff}}.
  }
\label{fig:mass}
\centerline{\psfig{figure=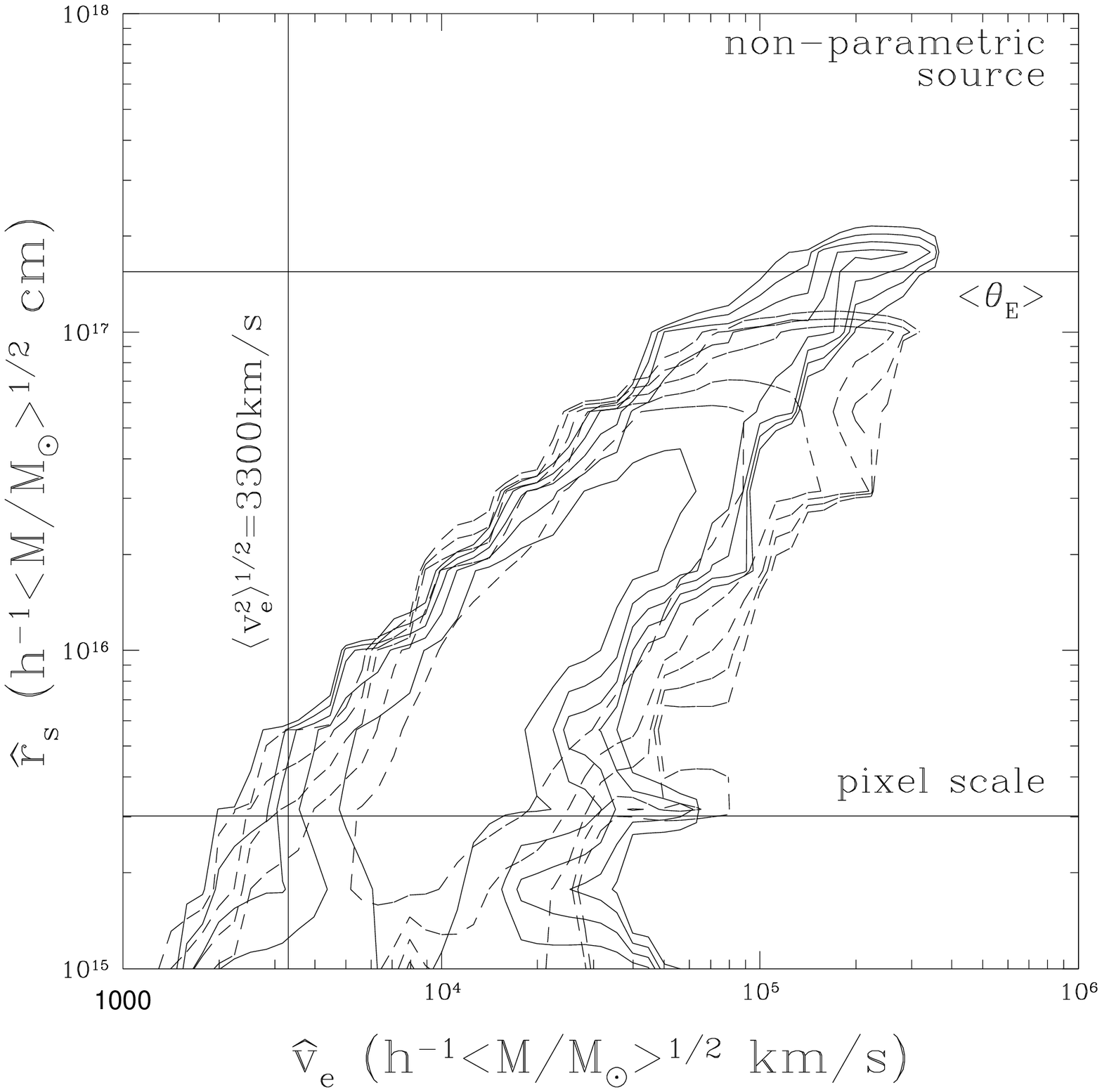,width=3.20in}}
\caption{ Likelihood contours for the effective source velocity, $\hat{v}_e$, and the
  scaled source size, $\hat{r}_s$ in the non-parametric models.
  The solid (dashed) contours are for the Gaussian
  (thin disk) source model.  Contours are drawn at intervals of $\Delta\log(L/L_{max})=1$. 
  The horizontal lines show the scales for $\hat{r}_s$ corresponding to the Einstein
  radius $\langle\theta_E\rangle$ of the average mass star and the pixel scale of the
  magnification maps.  The vertical line shows our estimate of 
  $\langle v_e^2\rangle^{1/2}=3300$~km/s for the rms source plane velocity.
  }
\label{fig:rscont}
\end{figure}

We analyzed the data using both parametric and non-parametric treatments for
the variability of the source.  For the parametric models we assumed
a constant source with $\sigma_0=0.05$~mag of additional variability,
separately modeling the individual images (Eqn.~\ref{eqn:chi1}).  
For each set of physical parameters ${\bf \xi_p}$ we tested $3\times 10^6$, 
$5\times 10^5$, $10^8$, and $3\times 10^6$ trajectories for images A, B, C 
and D respectively.  The number of trials was set so that of order $10^4$
trial trajectories would pass a threshold of $\chi^2 < 3N_{dof}$ for
cases with reasonable physical parameters.  The number of trials was
highest (lowest) for image C (B) because it has the most (least) complex
light curve (see Fig.~\ref{fig:data2237}).  For the non-parametric models we 
fit all four images simultaneously (Eqns.~\ref{eqn:chiall1} and \ref{eqn:chiall2})
using $10^8$ trial light curves for each set of physical parameters. The
threshold of $\chi^2_{max}=5N_{dof}$ was set to get approximately $10^3$
trial trajectories past the threshold for each set of physical parameters.
As in any Bayesian approach, only the relative probabilities of the physical 
parameters are estimated, so the absolute numbers of trials and the differences 
in the number of trials for the images has no effect on the results.  We 
performed all the calculations on two independent realizations of the
magnification patterns for each image and stellar mass fraction
to check that the $40\langle\theta_E\rangle$ regions
were large enough to provide a fair sample of light curves and that 
the probability estimates had converged.    We found
no significant differences between the results for the independent realizations
and discuss only the combined results.    We focus on the
results for the non-parametric models because they avoid the assumptions about 
source variability required by the parametric models.  In general, however,
the two approaches give consistent results for all physical
variables given their uncertainties.

\subsection{ The Effective Source Velocity and the Average Stellar Mass \label{sec:ve} }

The effective source velocity $\hat{v}_e$ determines the time scale for observing
microlensing events, and can be used to estimate the average mass $\avgm$ of
the microlenses given a prior probability distribution for the true source 
velocity $v_e$ (Eqn.\ref{eqn:massest}).  
Fig.~\ref{fig:veff} shows our estimate of the effective source velocity $\hat{v}_e$
after marginalizing over all other variables based on the parametric and
non-parametric analysis methods.
The parametric model gives a median velocity estimate of $\hat{v}_e=39000$~km/s with a 
68\% confidence region of $21600~\hbox{km/s} \ltorder \hat{v}_e \ltorder 71200~\hbox{km/s}$,
while the non-parametric model gives a median of $\hat{v}_e=19800$~km/s with a 
68\% confidence region of $10200~\hbox{km/s} \ltorder \hat{v}_e \ltorder 39600~\hbox{km/s}$.  
While the two estimates are statistically consistent, the differences have significant 
implications for estimates of $\avgm \propto \hat{v}_e^{-2}$.  
The non-parametric models generally find
intrinsic fluctuations in the source that have significant, slow temporal variations 
that will not be well-modeled by the assumed constant source (plus $\sigma_0=0.05$~mag 
fluctuations) used in the parametric models (see \S\ref{sec:lcurves}).   
Thus, a likely hypothesis for the origin
of the differences in the velocity estimates is that the parametric models are forced
to create some of the variability which is actually intrinsic to the source using
microlensing, and this is most easily done by increasing the effective velocity and
the source size.

The parametric model, where the light
curves of each image were evaluated separately, also gives probability distributions
for the velocity for the individual images, as also shown in Fig.~\ref{fig:veff}.  While
the four images give mutually consistent estimates of the effective velocity,
the two images with strong features in the light curve (A and C, see Fig.~\ref{fig:data2237})
dominate the results.  Image B, whose light curve is dominated by a slow drift,
favors slower velocities as this makes it more likely to avoid having features.  Image
D has a bimodal velocity distribution produced by two different regimes for the
size of the source.  When the source is small, the light curves can be reproduced
using velocities similar to images A and C.  However, there is a higher likelihood
region where the source size is large and the effective velocity is very high.
This solution branch is similar to that proposed by Refsdal \& Stabell~(\cite{Refsdal93}),
where a heavily smoothed magnification pattern makes it easy to reproduce the broad,
low amplitude peaks in the D light curve but requires a very high effective
velocity because the smoothing also increases the scale length of the variations
in the magnification pattern.

Our estimate of $\langle v_e^2\rangle^{1/2}=3300$~km/s for the typical source velocity 
is significantly lower than
the effective velocity $\hat{v}_e$ estimated from fitting the light curves.  This means that
the average mass of the microlenses must be significantly less than solar.  
Fig.~\ref{fig:mass} shows the estimate of $\avgm$ found by convolving the 
two velocity estimates as a function of the mass (Eqn.~\ref{eqn:massest}).  
The parametric models, because of their very high estimated of $\hat{v}_e$,
give very low mass estimates. The median estimate of the
mass is $\avgm=0.016 h^2 M_\odot$ with a 68\% (90\%) confidence range of
$0.0015 h^2 M_\odot < \avgm < 0.16 h^2 M_\odot$ 
($0.00032 h^2 M_\odot < \avgm < 0.88 h^2 M_\odot$).  The non-parametric
models, because of their lower estimates of $\hat{v}_e$, give higher mass
estimates.   The median estimate of the
mass is $\avgm=0.037 h^2 M_\odot$ with a 68\% (90\%) confidence range of
$0.0059 h^2 M_\odot < \avgm < 0.20 h^2 M_\odot$
($0.0015 h^2 M_\odot < \avgm < 0.56 h^2 M_\odot$).  There are roughly
equal contributions to the uncertainties from the estimate
of the effective source velocity in our fits and the estimate of the true 
source velocity.  Unfortunately, the mass scale depends on the square of the
velocity, so the errors on the estimate of the mass scale are substantial. 
We may also have inadvertently biased the mass scales downwards by restricting
our analysis to the OGLE light curves.  The variability of the quasar during
this period was significantly greater than during the preceding decade (see
Corrigan et al.~\cite{Corrigan91}, Ostensen et al.~\cite{Ostensen96} and
Wozniak et al.~\cite{Wozniak00a}, \cite{Wozniak00b}),
so expanding our analysis to the earlier data would probably lower the estimate
of the effective velocity.

\begin{figure}
\centerline{\psfig{figure=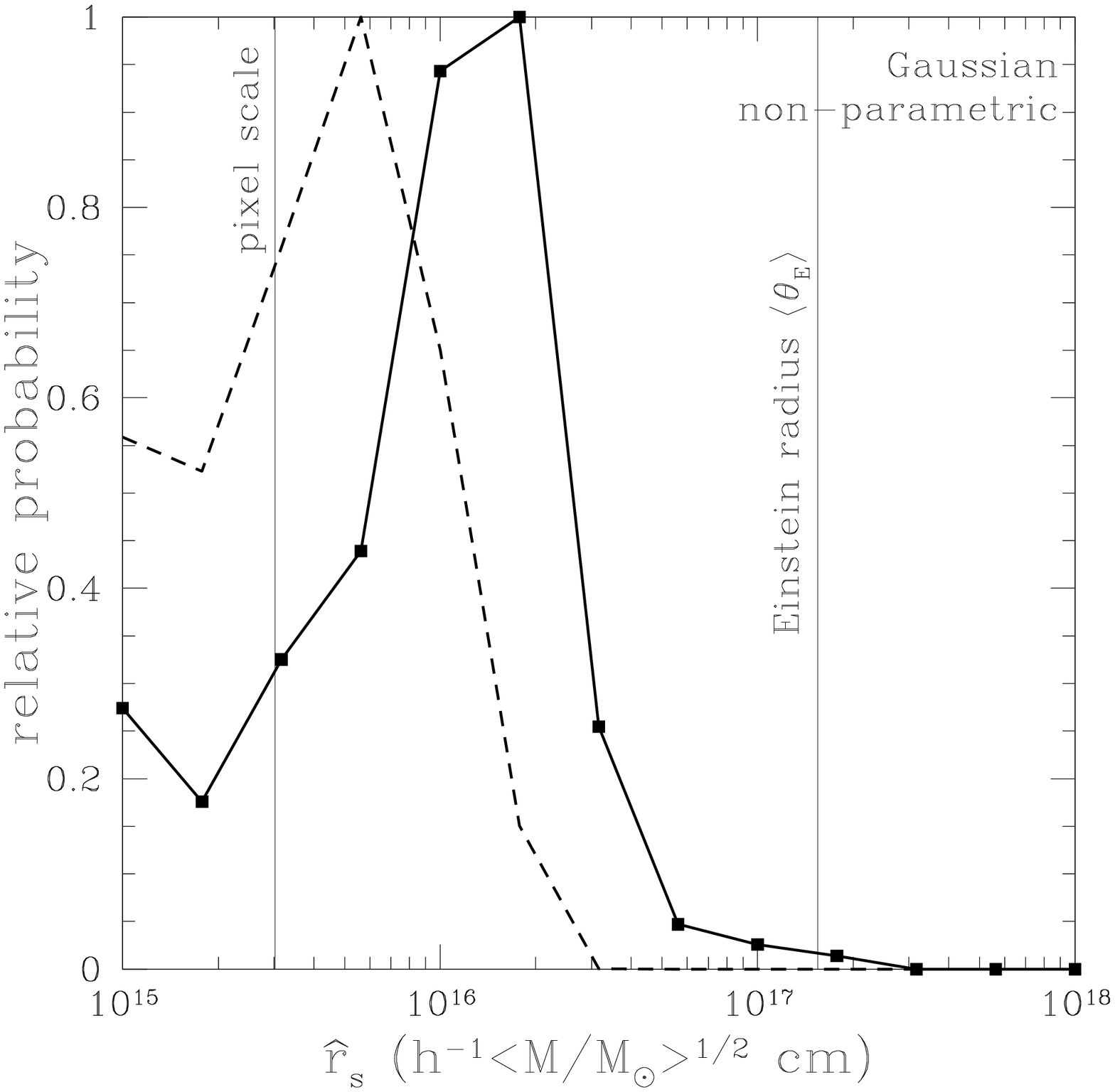,width=3.25in}
            \psfig{figure=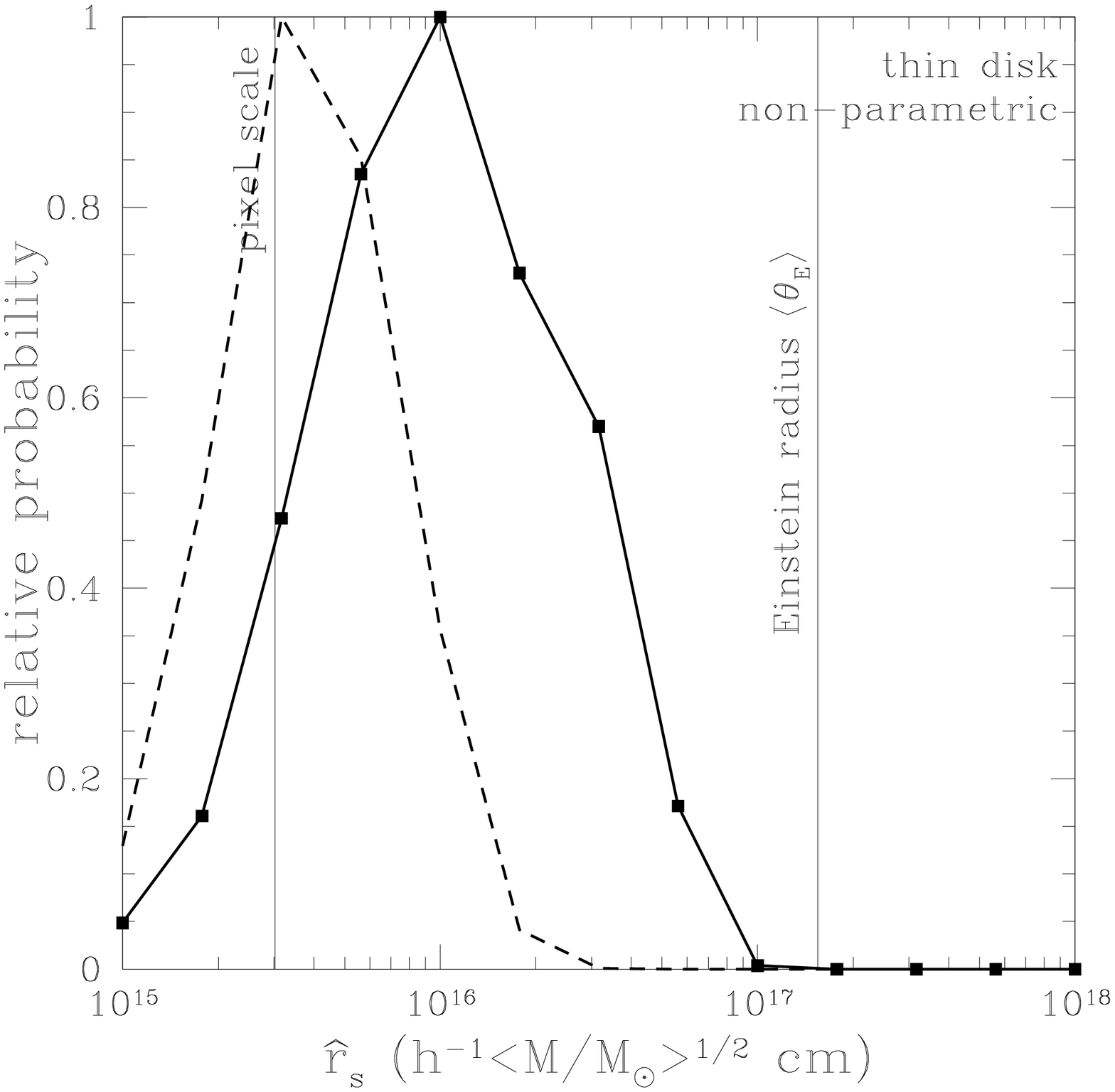,width=3.25in}}
\caption{ Probability distributions for the scaled source size $\hat{r}_s$
  in the non-parametric models and either the Gaussian (left) or thin disk (right)
  model for the disk surface brightness profile.  
  The heavy dashed line shows the estimate for $\hat{r}_s$ with 
  a prior of $0.2h^2 M_\odot < \langle M\rangle < 2 h^2 M_\odot$ on
  the mass of the stars.  The two vertical
  lines show the Einstein radius 
  $\langle\theta_E\rangle$ corresponding to a star with the average mass
  $\langle M\rangle$ and the pixel scale of the magnification maps.
  }
\label{fig:rsrc}
\centerline{\psfig{figure=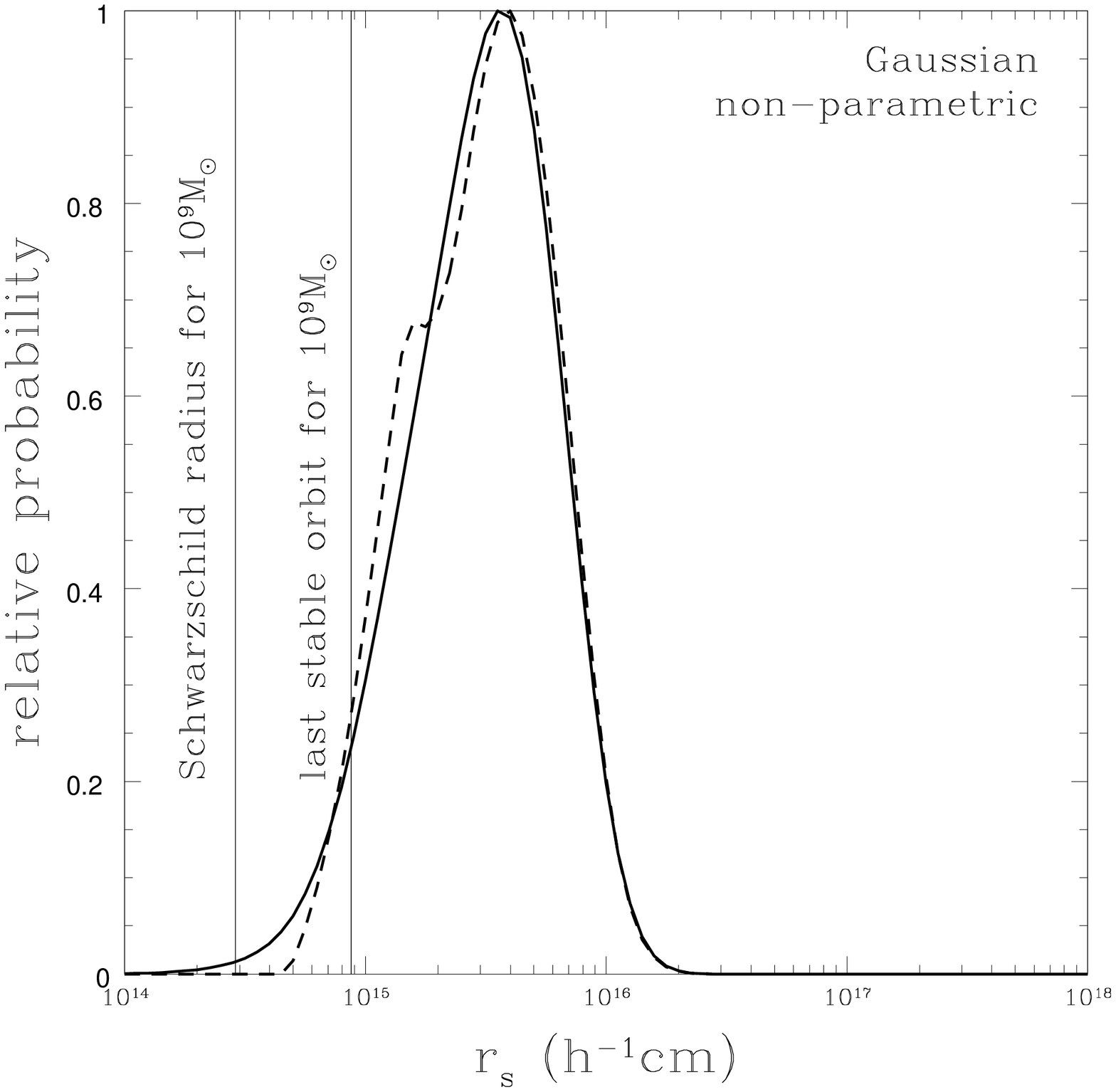,width=3.25in}
            \psfig{figure=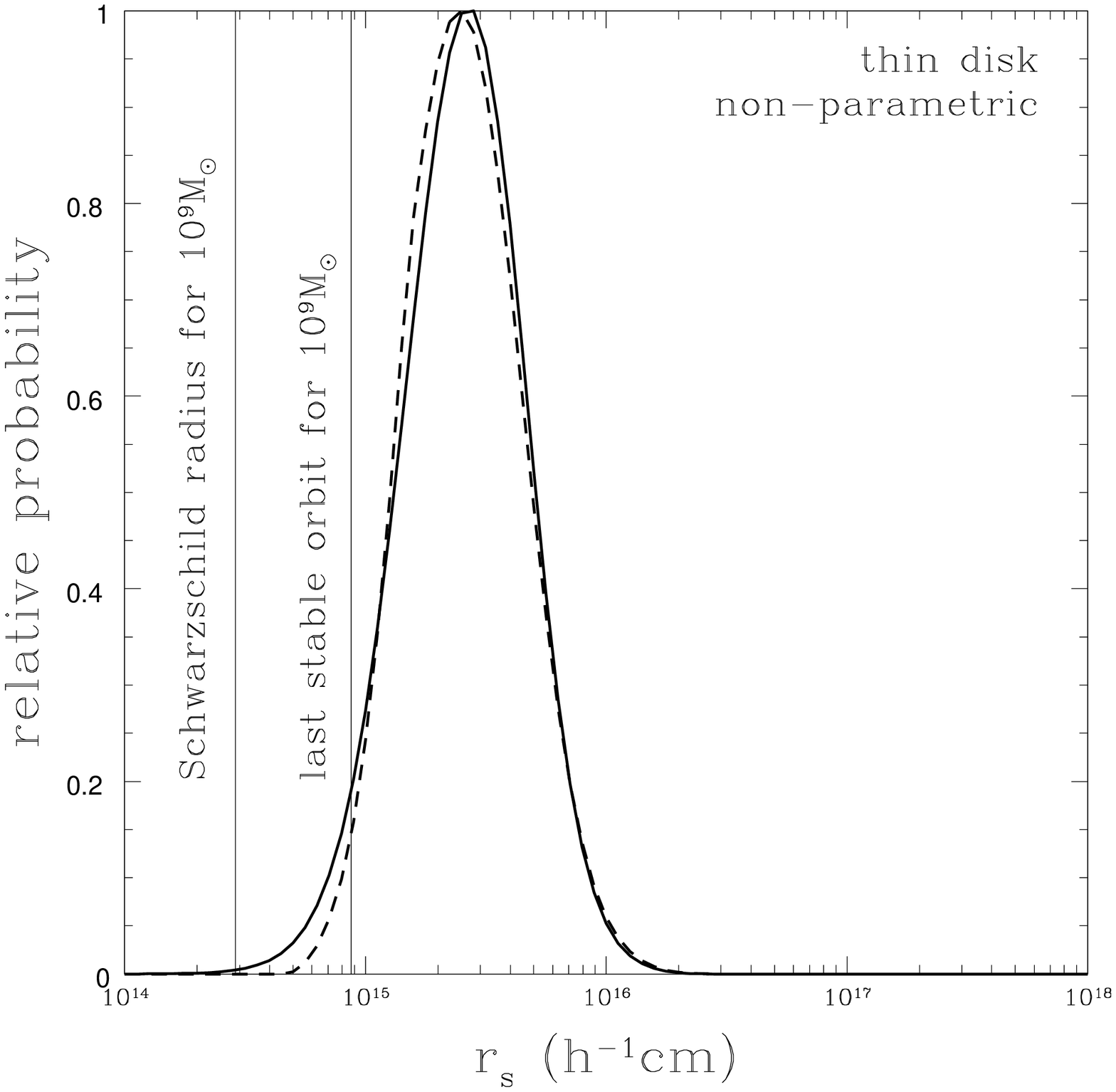,width=3.25in}}
\caption{ Probability distributions for the physical source size $r_s$ in the
  non-parametric models for the Gaussian (left) and thin disk (right) models
  for the disk surface brightness.
  The dashed curve shows the estimate for $r_s$ with
  a prior of $0.2h^2 M_\odot < \langle M\rangle < 2 h^2 M_\odot$ on
  the mass of the stars.
  The vertical line shows the Schwarzschild radius $R_{BH}$ of a $10^9 M_\odot$
  black hole.  The last stable orbit lies at $3R_{BH}$.  The results from the
  parametric models are identical.
  }
\label{fig:rsrctrue}
\end{figure}

\subsection{The Scaled Source Size \label{sec:rshat} }

The source structure and the scaled source size $\hat{r}_s$ control the smoothing
of the magnification pattern, and the amount of smoothing has a powerful effect
on the effective velocity.  Fig.~\ref{fig:rscont} shows likelihood contours for
$\hat{r}_s$ and $\hat{v}_e$ for both source structures in the non-parametric 
models.  There is a strong, essentially linear correlation between the two 
variables in the sense that larger sources require higher velocities,
with $\log(\hat{r}_s/h^{-1}\hbox{cm}) \simeq 15.8 + \log(\hat{v}_e/10^4\hbox{km/s})$. 
While the main ridge in the likelihood is similar for both analysis methods,
the parametric models have a more extended tail of high velocity solutions
as discussed in \S\ref{sec:ve}.  The region of acceptable solutions 
extends to regions with more compact sources than can be resolved by our
standard magnification maps, so our lower limits on $\hat{r}_s$ are
unreliable.  This was a consequence of the trade off between
high resolution magnification maps and magnification maps containing large
numbers of statistically differing regions. 

When we marginalize the likelihoods over the velocity, we find the estimates of
the source size and structure shown in Fig.~\ref{fig:rsrc}.  The thin disk model
is favored over the Gaussian model in both analysis methods, with the probability
of the thin disk model being 96\% for the parametric analysis and 76\% for the
non-parametric analysis.  While the probability distributions for the source 
size are statistically consistent, the parametric models favor larger sources
than the non-parametric models.  For the Gaussian source we find 68\% 
confidence regions of 
$8.0 \times 10^{15} h^{-1} \hbox{cm} \ltorder \hat{r}_s \ltorder 3.6 \times 10^{16} h^{-1} \hbox{cm}$
and
$3.5 \times 10^{15} h^{-1} \hbox{cm} \ltorder \hat{r}_s \ltorder 2.4 \times 10^{16} h^{-1} \hbox{cm}$
for the parametric and non-parametric methods.  For the thin disk 
models we find 68\% confidence regions of
$1.1 \times 10^{16} h^{-1} \hbox{cm} \ltorder \hat{r}_s \ltorder 5.7 \times 10^{16} h^{-1} \hbox{cm}$
and
$4.1 \times 10^{15} h^{-1} \hbox{cm} \ltorder \hat{r}_s \ltorder 2.6 \times 10^{16} h^{-1} \hbox{cm}$
for the parametric and non-parametric methods.  The shifts in the distributions for
$\hat{r}_s$ simply match the shifts in the estimates of $\hat{v}_e$ because of the 
strong correlation of these two variables (Fig.~\ref{fig:rscont}).  
The peaks of the probability 
distributions correspond to scales that are well-resolved in our magnification
maps ($\log\hat{r}_s=16$ corresponds to 3.3~pixels, so the source averages the
magnification pattern over roughly $\pi \hat{r}_s^2=35$ pixels).  The distributions
decrease significantly before reaching the pixel scale, but it is clear that there
are significant tails to the distribution that we have not fully resolved.

We also explored the consequences of imposing a prior of
$0.2 h^2 M_\odot < \avgm < 2 h^2 M_\odot$.
on the mass of the microlenses.  Forcing a higher mass with a fixed
source velocity $v_e$ rules out solutions with high effective velocities
$\hat{v}_e$ and large source sizes $\hat{r}_s$ (see  Fig.~\ref{fig:rsrc}).  
The 68\% confidence regions for the Gaussian source become
$4.2 \times 10^{15} h^{-1} \hbox{cm} \ltorder \hat{r}_s \ltorder 1.4 \times 10^{16} h^{-1} \hbox{cm}$
and
$1.9 \times 10^{15} h^{-1} \hbox{cm} \ltorder \hat{r}_s \ltorder 1.1 \times 10^{16} h^{-1} \hbox{cm}$
for the parametric and non-parametric methods, and they become
$3.7 \times 10^{15} h^{-1} \hbox{cm} \ltorder \hat{r}_s \ltorder 8.9 \times 10^{15} h^{-1} \hbox{cm}$
and
$2.1 \times 10^{15} h^{-1} \hbox{cm} \ltorder \hat{r}_s \ltorder 8.0 \times 10^{15} h^{-1} \hbox{cm}$
for the thin disk model and the parametric and non-parametric methods.  The lower
limits in this case are significantly affected by the pixel scale of the 
magnification maps.

\subsection{The Structure of the Accretion Disk and the Mass of the Black Hole 
  \label{sec:rs}}

We can measure the physical source size of the disk, $r_s$, more accurately
than the scaled source size, $\hat{r}_s$, because of the nearly linear correlation between
$\hat{v}_e$ and $\hat{r}_s$ ($\hat{r}_s \propto \hat{v}_e^x$ with $x\simeq 1$,
\S\ref{sec:rshat}, Fig.~\ref{fig:rscont}).  Since $r_s = \hat{r}_s \avgm^{1/2}$
and $\avgm  \propto (v_e/\hat{v}_e)^2$, the physical size of the source 
$r_s \propto v_e \hat{v}_e^{x-1} \simeq v_e $ depends on our estimate of
the physical velocity $v_e$ but avoids the degeneracies between $\avgm$,
$\hat{v}_e$ and $\hat{r}_s$.  This is illustrated in 
Fig.~\ref{fig:rsrctrue}, where we see that the estimates of $r_s$ 
are unaffected by the addition of the prior on $\avgm$.  They are
also independent of the statistical method even though the scaled
source radii are larger in the parametric models. 
Adopting the non-parametric source without a prior to be the fiducial
case, we find that the median estimate for the Gaussian source size is
$r_s=3.6  \times 10^{15} h^{-1}\hbox{cm}$  
($1.6\times 10^{15}h^{-1}\hbox{cm}\ltorder r_s \ltorder 6.9 \times 10^{15}h^{-1}\hbox{cm}$
at 68\% confidence) and that the median estimate for the thin disk
source size is
$r_s=2.9  \times 10^{15} h^{-1}\hbox{cm}$  
($1.6\times 10^{15}h^{-1}\hbox{cm}\ltorder r_s \ltorder 7.6 \times 10^{15}h^{-1}\hbox{cm}$
at 68\% confidence).

Because the thin disk model is a self-consistent, physical model for the accretion
disk, we can compute the disk luminosity from our estimate of the scale length
$r_s$.  Integrating over the surface brightness profile, we
find that the effective isotropic rest-frame luminosity of the (face-on) disk is  
\begin{equation}
    L_{V,model} =  { 16 \pi^2 C_{BB} r_s^2 h_P c^2 \Delta\lambda \over \lambda^5 }
               =  (2 \times 10^{45}) h^{-2} 
                  \left( {r_s \over 10^{15} h^{-1}\hbox{cm} } \right)^2 
                  \hbox{ergs/s}
\end{equation}
where $C_{BB}=2.58=\int_0^\infty x dx(\exp(-x^{3/4})-1)^{-1}$, 
$\Delta\lambda=827\hbox{\AA}/(1+z_s)\simeq 300\hbox{\AA}$
is the redshifted width of the V-band filter, 
$\lambda=5505\hbox{\AA}/(1+z_s)\simeq 2000\hbox{\AA}$ 
is the redshifted center of the V-band filter, and $h_P$ is Planck's constant.  
We can compare this estimate to the observed luminosity of the source after
correcting for magnification.  If the intrinsic source magnitude is $V_0$,
then the observed luminosity is
\begin{equation}
   L_{V,obs} = (6.2 \times 10^{45}) h^{-2} 10^{0.4(V_0-19)} \hbox{ergs/s}
\end{equation}
For $V_0=19\pm0.5$~mag, we need $r_s \simeq (1.7 \pm0.4) \times 10^{15}h^{-1}$~cm,
which is consistent with our direct estimate of the source size.
At least at this wavelength,
an optically thick, thermally emitting disk structure is consistent with the data.
Although the CIII] emission line lies in the V band, its equivalent width is
too small compared to the total width of the bandpass to significantly
modify these conclusions. 

\begin{figure}[t]
\centerline{\psfig{figure=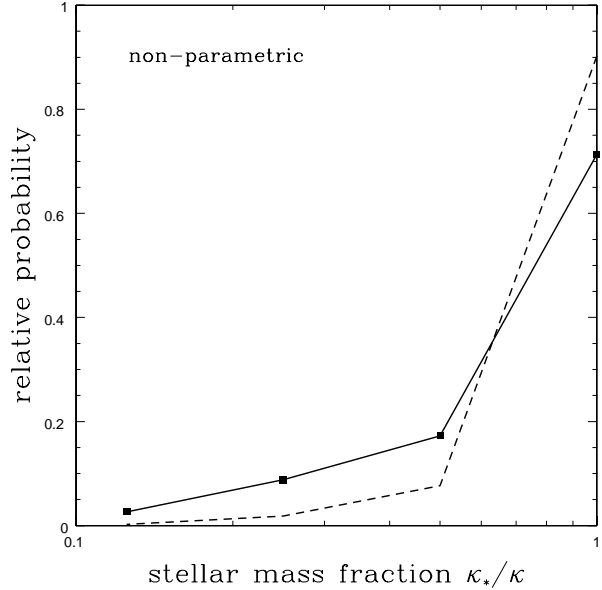,width=3.25in}}
\caption{ Probability distributions for the stellar mass fraction 
  $\kappa_*/\kappa$ in the non-parametric models.  
 The solid (dashed) curves show the probability distributions 
  for $\kappa/\kappa_*$ without (with) the strong mass prior.  }
\label{fig:massfunc}
\end{figure}

We can also use the thin disk model to infer the mass of the black hole given
that the temperature at radius $r_s$ is $T_s(r_s) \simeq 70000$~K. 
If all the viscous energy released is radiated 
locally, and we are well outside the Schwarzschild radius, then 
$\sigma T_s^4 = 3 G M_{BH} \dot{M}/8\pi r_s^3$, and the black hole mass is 
\begin{equation}
    M_{BH} \simeq 2.6 \times 10^8 \eta_{0.1}^{1/2} 
         \left( {r_s \over 10^{15} \hbox{cm} } \right)^{3/2}
         \left( { L \over L_E } \right)^{-1/2}
\end{equation}
where $\eta=0.1\eta_{0.1}$ is the overall efficiency of the accretion and $L/L_E$ is
the total luminosity in units of the Eddington luminosity.  Given our
estimate of $r_s$, this implies 
$M_{BH} \simeq 1.1 \times 10^9 h^{-3/2} M_\odot \eta_{0.1}^{1/2}(L/L_E)^{-1/2}$
($0.43  \times 10^9 M_\odot \ltorder M_{BH} \ltorder 2.5 \times 10^9 M_\odot$)
that the Schwarzschild radius is $R_{BH} \simeq 3.1 \times 10^{14} h^{-3/2}$~cm, and that
$r_s$ is approximately $ 8 h^{1/2}$ Schwarzschild radii.  For comparison, if we
estimate the mass from the V-band luminosity, we find 
$M \simeq (5 \pm 3) \times 10^9 M_\odot h^{-2}(0.01/f)(L/L_E)^{-1/2}$ 
where $f\sim 0.01$ is
the fraction of the radiation emitted in the V-band.   Thus, our derived 
structure for the accretion disk is roughly consistent with
the theory from which it is derived and the observed luminosity. There,
are however, some limitations.  First, we neglected the corrections to the
temperature profile near the last stable orbit (see \S\ref{sec:method}).  
Second, our thin disk model 
assumes a disk dominated by gas pressure and absorption opacity, both of which
have probably broken down on these scales and should be replaced by radiation 
pressure and scattering opacity.  Third, we assumed a face on disk, thereby neglecting
inclination effects.  Nonetheless, the self-consistency of the results is
reassuring.

\begin{figure}
\centerline{\psfig{figure=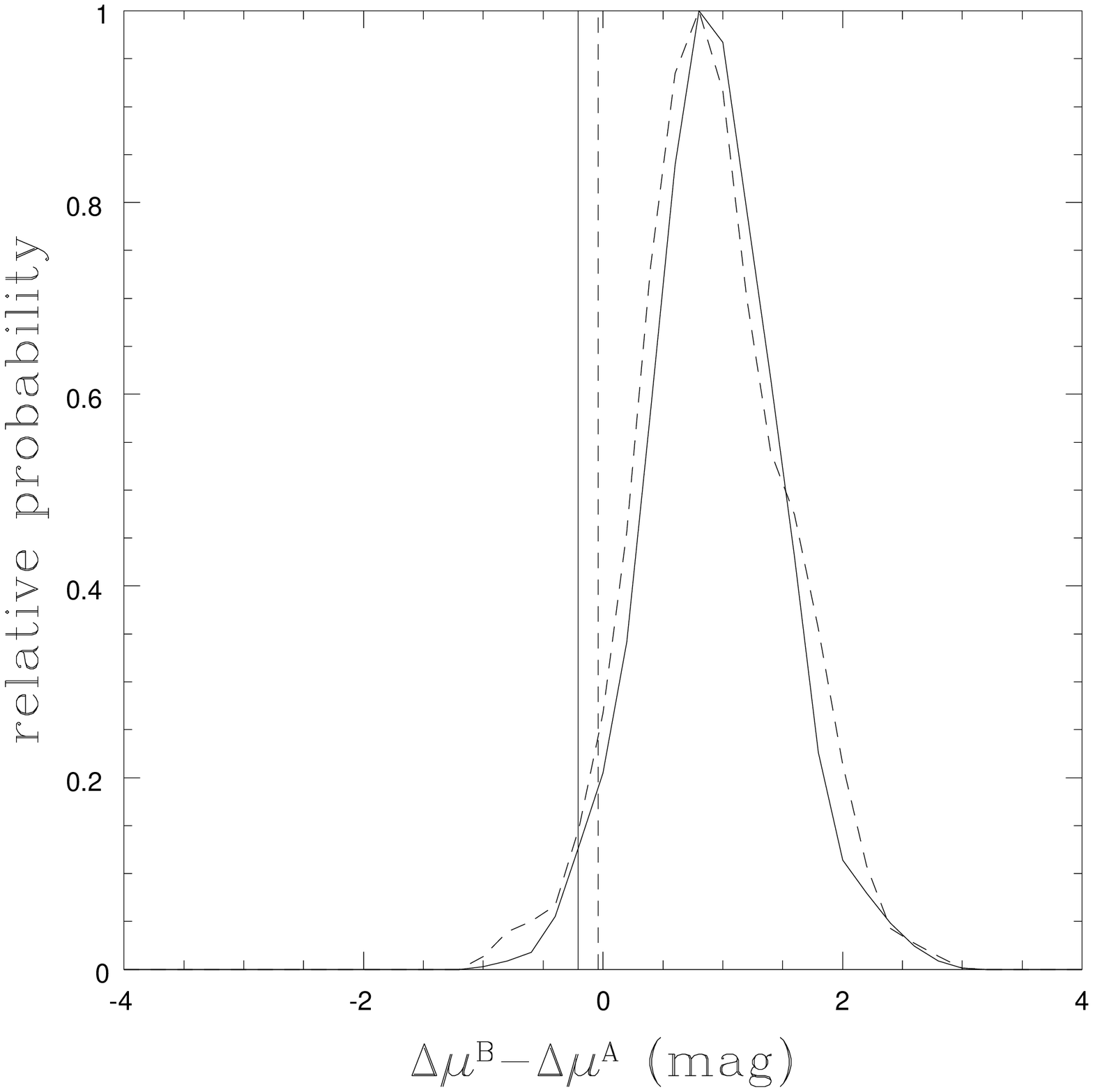,width=3.25in}
            \psfig{figure=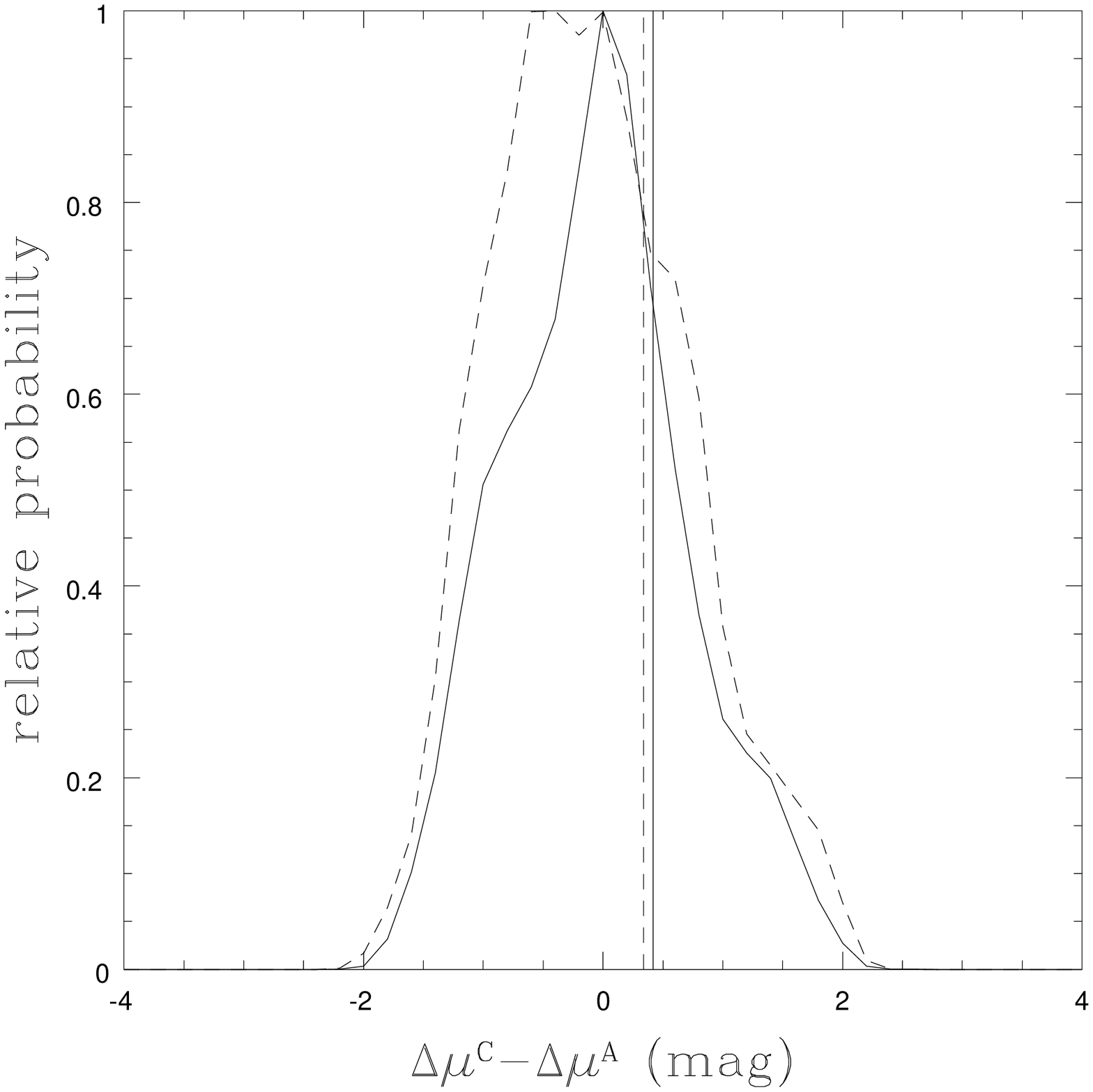,width=3.25in}}
\centerline{\psfig{figure=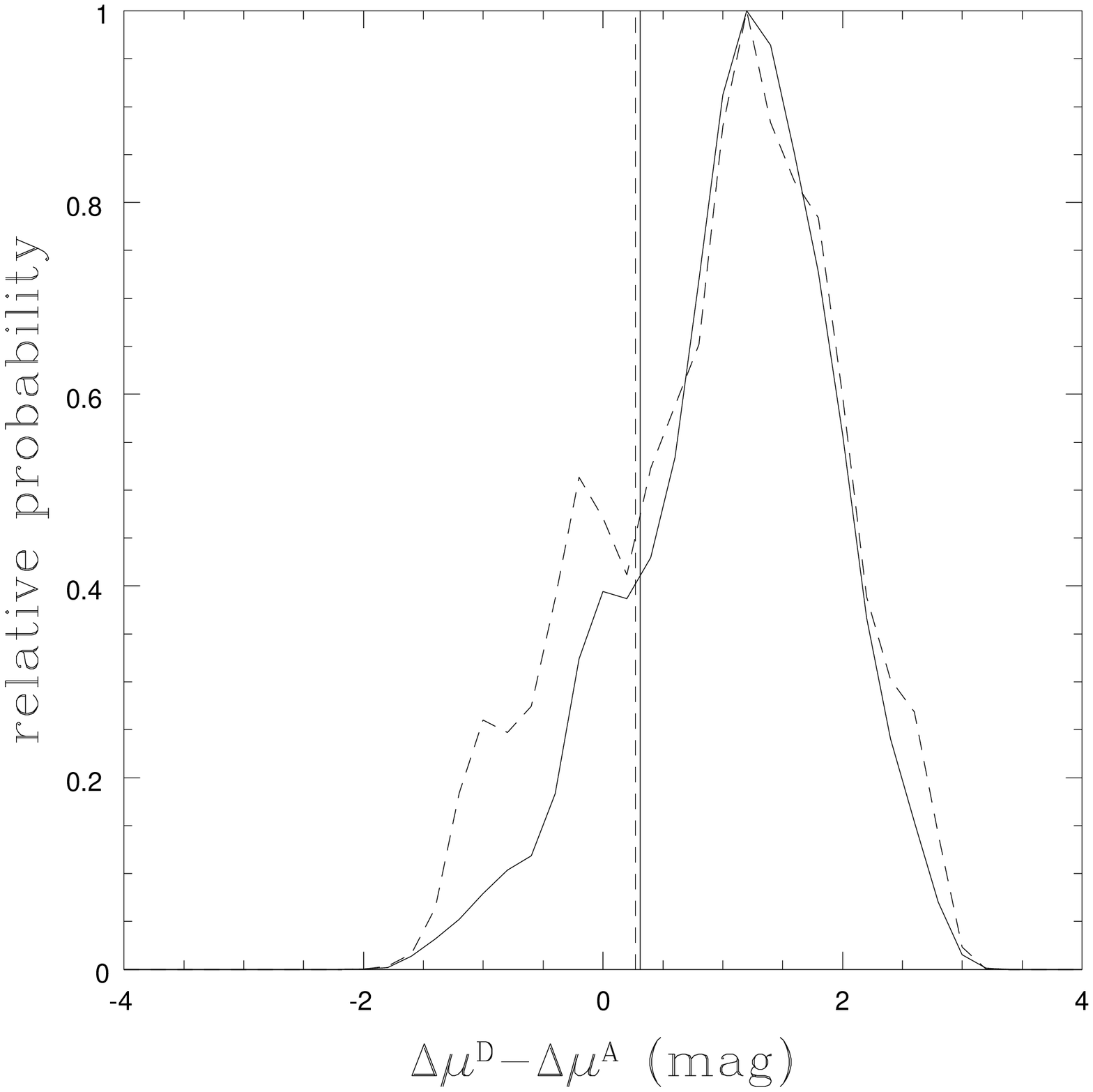,width=3.25in}}
\caption{ Probability distributions for offsets to the source magnitudes
  $\Delta\mu^\alpha$ relative to image A.   The dashed lines show the
  effect of imposing the strong mass prior.  The solid (dashed) vertical
  lines show the differential extinction estimates of Falco et al.~(\protect{\cite{Falco99}})
  (Agol et al.~\protect{\cite{Agol00}}). 
  }
\label{fig:fluxrat}
\end{figure}

\subsection{The Surface Density of Stars \label{sec:kappa} }

We find that the present models cannot distinguish between our two models
for the stellar mass functions as the
relative probabilities of the Salpeter ($x=2.35$, $r=100$) and mono-mass
($r=1$) mass functions are almost exactly equal.  This matches the general
conclusion from previous studies that it is difficult to recognize the
differences in the microlensing effects created by changing the mass 
function (see Paczynski~\cite{Paczynski86}, Wyithe et al.~\cite{Wyithe00b}).  
However, we do obtain estimates for the stellar
mass fraction, as shown in Fig.~\ref{fig:massfunc}.  For the parametric
(non-parametric) models the one-sided 68\% confidence limit is
$\kappa_*/\kappa > 0.28$ ($0.52$).  The difference is again due to
the shift in the permitted range for $\hat{v}_e$ between the two analysis
methods. With fewer stars the source must have a higher velocity
to keep a fixed level of photometric variability, so the lower
stellar fraction models are more viable in the parametric models. 
Imposing the $0.2h^2 M_\odot < \avgm < 2.0 h^2M_\odot$ prior on the
mass of the microlenses leads to much stronger bounds on the stellar
surface density of $\kappa_*/\kappa > 0.40$ ($0.73$) for the same 
reason -- the mass prior forces a lower effective velocity which favors
higher stellar mass fractions.  Given that 
the images pass through the central regions of the bulge of a nearby spiral 
galaxy, we would expect the surface density to be dominated by the stars.

We did not consider changes in the total surface density of the lens, but
we can estimate the consequences of changes in the macro model by using
the generalized versions of the mass sheet degeneracy (Paczynski~\cite{Paczynski86}
for the case of microlensing) discussed in Appendix~\ref{sec:rayshoot}.  We used models
with fixed total surface density $\kappa=\kappa_s+\kappa_*$ and a range
for the fraction $\kappa_*/\kappa$ composed of stars.  Each of these models
is equivalent to a model with no smoothly distributed dark matter and
$\kappa'=\kappa_*'=\kappa_*/(1-\kappa_s)$.  For example,
the models of image A with $\kappa=0.394$ and $\kappa_*/\kappa=1$, $1/2$, $1/4$ 
and $1/8$ are the same as models with $\kappa_s'=0$ and 
$\kappa'=\kappa_*'=0.394$, $0.245$, $0.140$ and $0.075$ respectively.
Thus, the model sequence in $\kappa_*/\kappa$ is related to macro model
sequence with $\kappa=\kappa_*$ and an increasingly concentrated
mass distribution.  It does not quantitatively match any real macro
model sequence because the 4 images must be scaled independently.
We can keep the source plane
length and velocity scales fixed ($\beta=1$) by increasing the 
microlens mass scale, $\avgm' = \avgm/(1-\kappa_s)^{1/2}$.  Hence,
the models $\kappa_*/\kappa<1$ models when rescaled to have 
$\kappa'=\kappa_*'$  would be less affected by the mass prior.
Nonetheless, these scaling arguments suggest that the OGLE light
curves would tend to rule out mass distributions more centrally
concentrated than our standard isothermal model.

\subsection{The Flux Ratios of the Images \label{sec:fluxrat}}

In these models we have solved for the optimal magnitude shifts, $\Delta\mu^\alpha$,
between the observed image magnitudes and those expected from the source magnitude
and the macro model magnifications of $\mu^\alpha$ (in magnitudes, see \S\ref{sec:chidef}).  
If the light
curves correspond to a ``fair'' sample of the magnification patterns, then the 
magnitude shifts should converge to a model-independent value corresponding to
any error in the macro magnification or other systematic shifts such as differential
extinction between the images.  If the light curves are not a fair sample, then 
there will be a distribution of shifts depending on the location of each source
trajectory in the overall magnification pattern.  
The simplest means of estimating which light curve comes closest to matching the
mean magnification is to pick the light curve with the largest flux variations
compared to the range of magnifications in the magnification maps for that image.
For the raw magnification
maps (whose pixel scale corresponds to a source which is a little too small),
the dynamic ranges of the maps are approximately 60, 60, 300 and 200 for the
A, B, C and D images respectively, so we would expect either the A or B images
to come closest to converging to the mean magnification given the peak-to-peak
light curve amplitude ratios of $2.0$, $1.8$, $3.3$ and $1.9$ for the light curves.  Even
so, no light curve has sufficient dynamic range to have sampled the full range
of the magnification maps unless the source size is large.
Fig.~\ref{fig:fluxrat} shows
the probability distributions for $\Delta\mu^B-\Delta\mu^A$, $\Delta\mu^C-\Delta\mu^A$
and $\Delta\mu^D-\Delta\mu^A$ both with and without the strong mass prior.  These
were computed only for the non-parametric model of the source.

We can compare the values of the $\Delta\mu^\alpha$ to estimates of the
differential extinction between the images.  Agol et al.~(\cite{Agol00}) estimated
total extinctions from the color of the lens galaxy near each image to find
V-band differences of $-0.04\pm 0.29$, $0.42 \pm 0.37$ and $ 0.27 \pm 0.34$~mag
for the A, B and C images relative to image D.  Falco et al.~(\cite{Falco99})
estimated differential extinctions using the colors of the lensed images 
to find V-band differences of $-0.21 \pm 0.13$, $0.34 \pm 0.13$, and
$0.31 \pm 0.13$~mag for the B, C and D images relative to image A.  The two
sets of estimates are mutually consistent.  The differential extinction estimates 
have smaller uncertainties, but are more subject to systematic errors created
by microlensing.  If we add a term to the $\chi^2$ to force the offsets
to agree with the Falco et al.~(\cite{Falco99}) differential extinction
estimates with the uncertainties rounded upwards to $0.2$~mag, we can 
examine the effects of the offsets on all the other physical variables.  When
we do so, we find a weak effect towards suppressing models with larger values
of $\hat{r}_s$, but little else.    

\begin{figure}
\centerline{\psfig{figure=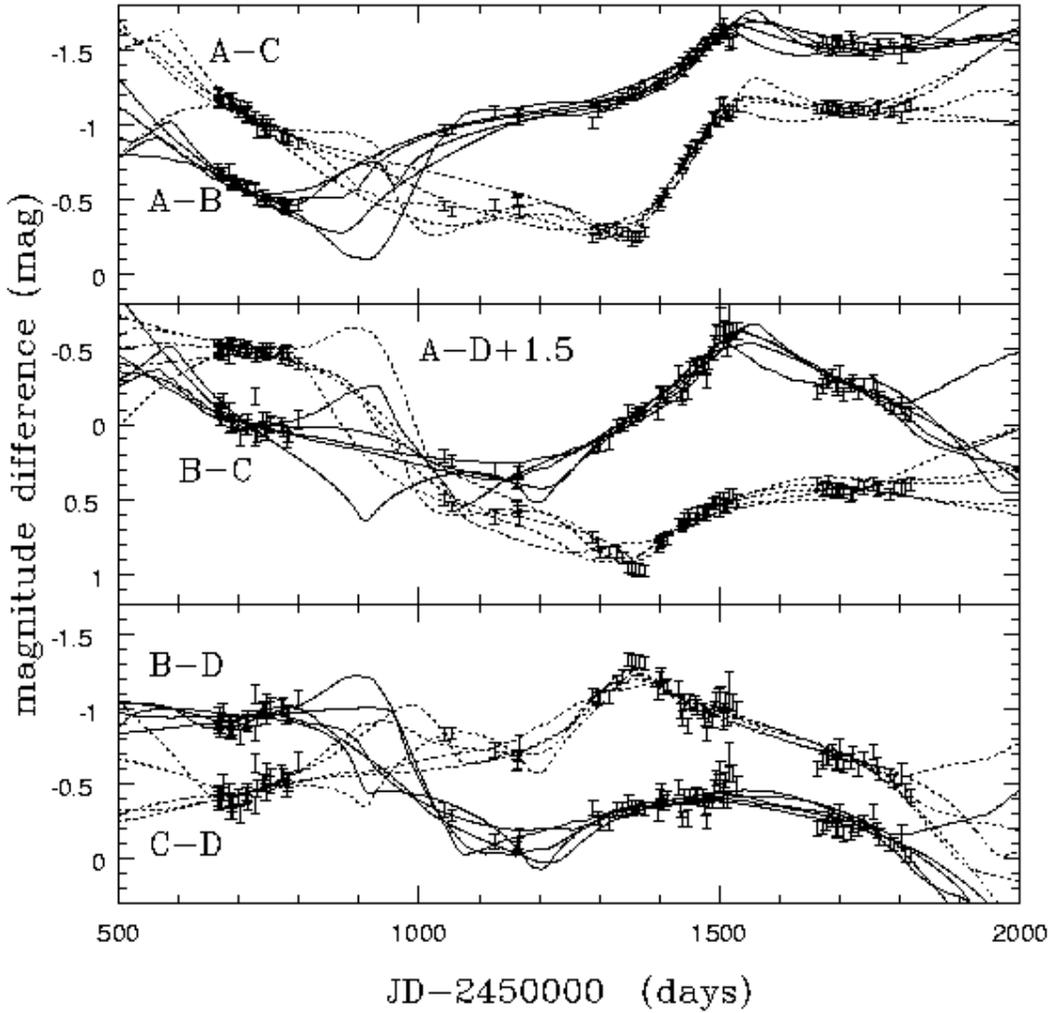,width=6.0in}}
\caption{ Difference light curves for the OGLE data and the 5 model light curves. 
  The points show the 6 possible difference light curves ($m_i^\alpha-m_i^\beta$ with
  $\alpha \neq \beta$) that can be constructed from the OGLE data.  The error bars
  are the OGLE uncertainties combined with a $\sigma_0=0.05$~mag systematic error in
  quadrature.  The curves show the model light curve differences.
  The vertical scale of each panel is $2.0$~mag.
  }
\label{fig:curve1}
\end{figure}

\begin{figure}
\centerline{\psfig{figure=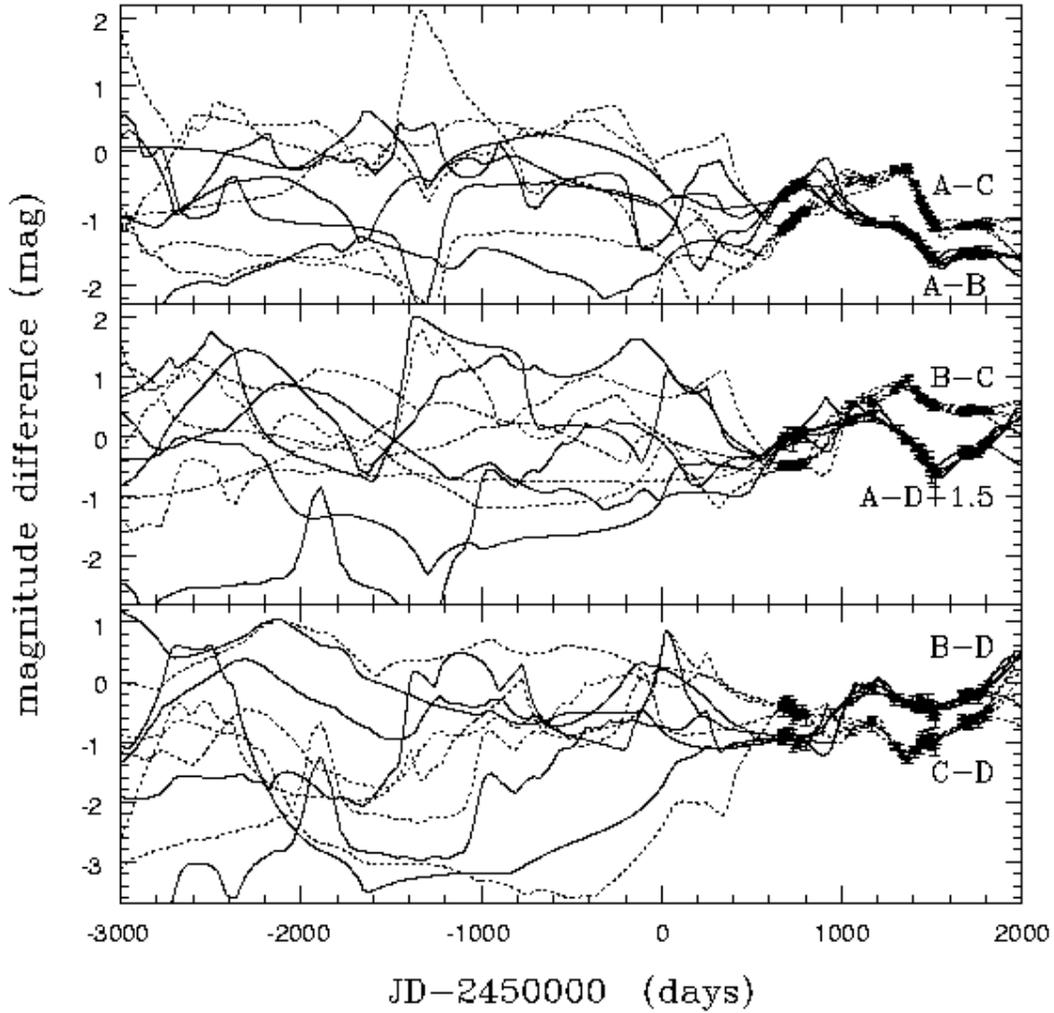,width=6.0in}}
\caption{ Difference light curves for the OGLE data and the 5 model light curves. 
  We show the same light curves as in Fig.~{\protect{\ref{fig:curve1}}},
  but with the time scale expanded to show the behavior of the model light 
  curves during the ten years before the start of the OGLE monitoring period.
    The vertical scale of each panel has been
  expanded to $5.0$~mag compared to the $2.0$~mag used in
  Fig.~{\protect\ref{fig:curve1}}.
  }
\label{fig:curve2}
\end{figure}

\begin{figure}
\centerline{\psfig{figure=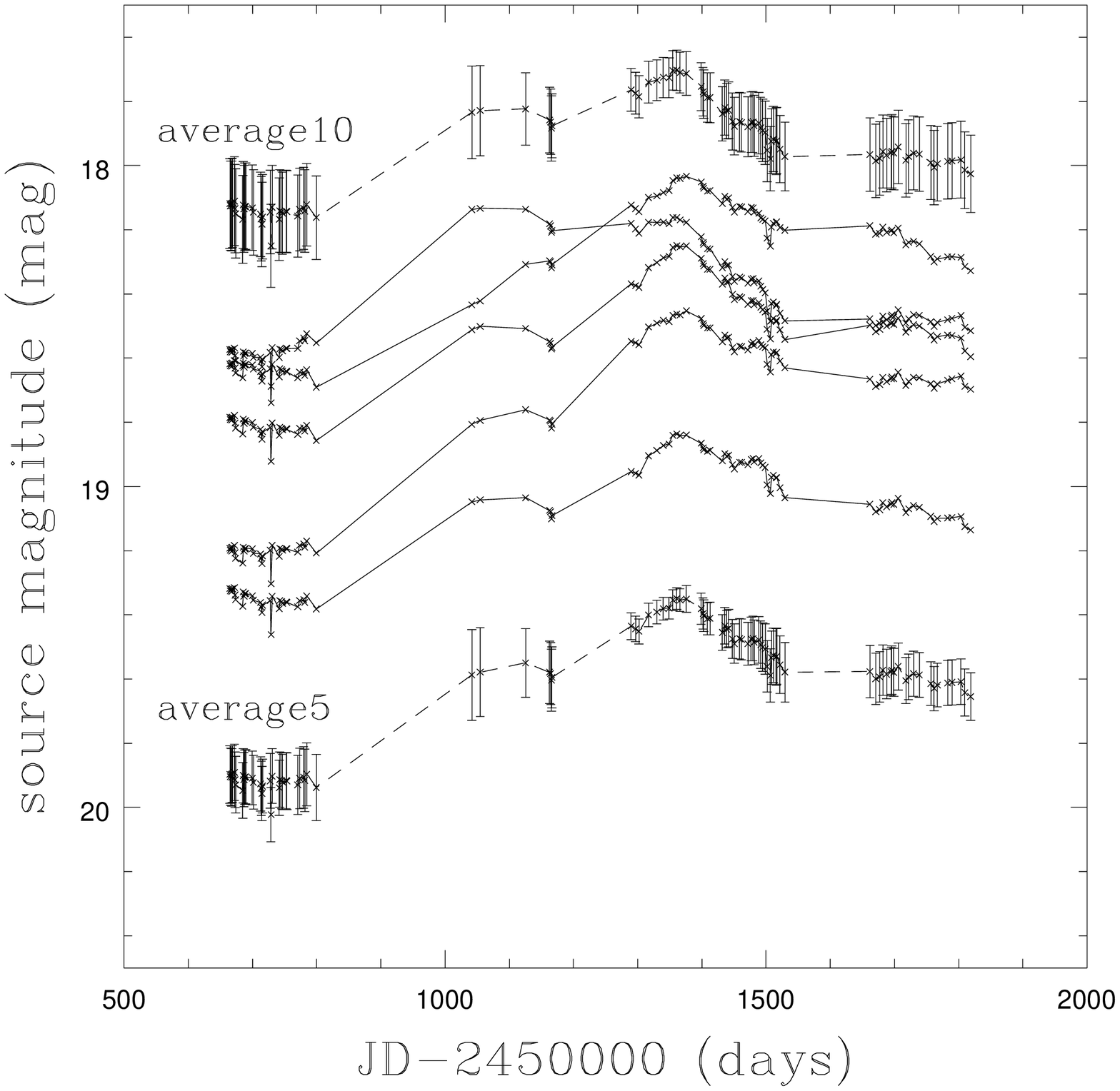,width=6.0in}}
\caption{ The reconstructed source fluxes for the light curves shown in 
  Fig.~{\protect\ref{fig:curve1}}.
  The lines connecting the points are only to guide the eye -- the source flux
  can be estimated only where there is data.  The shifts between the curves
  are another manifestation of how the monitoring period is not long enough for the light
  curves to determine the mean magnification (see \S{\protect\ref{sec:fluxrat}} and 
  Fig.~{\protect\ref{fig:fluxrat}}).
  The points connected by the dashed lines
  show the mean source light curves found after averaging either the
  5 source flux models shown here (``average5'') or these 5 plus the next 5 best 
  models (``average10'').  The
  mean magnitude of each source light curve was subtracted first, and 
  the error bars show the dispersion of the light curves.
  The mean magnitude of these averaged light curves is arbitrary and was
  set simply to keep the light curves well-separated.
  }
\label{fig:srccurve}
\end{figure}

\subsection{Examples of Light Curves \label{sec:lcurves} }

In this section we examine 5 of the 6 best light curve realizations found for
the non-parametric models.  We only save the light curves of the best model
found for each set of physical parameters after varying all the variables
for generating light curves (trajectory origin and velocity).  The fourth
best model had the same physical parameters as the third best, so its light
curve was not preserved.  
All 5 cases have $\chi^2\simeq 200$ compared to $N_{dof}=290$, slightly
over fitting the data given the additional $\sigma_0=0.05$~mag of systematic error
we added to each data point (i.e. if we reduced $\sigma_0$ to $0.04$~mag we 
would find $\chi^2 \simeq N_{dof}$).  Three of the cases have $\log\hat{r}_s=15.75$,
one has $\log\hat{r}_s=15.50$, and one has $\log\hat{r}_s=16.25$.  All have
$\kappa_*/\kappa=1$ and four out of five are thin disk models.  The effective
source velocities are $\hat{v}_e=14000$, $14600$, $29000$, $7100$ and $40000$~km/s
respectively.  

The goodness of fit of the non-parametric models is determined by how well the
model light curves reproduce the six possible light curve differences 
(Eqn.~\ref{eqn:chiall2}).  In Fig.~\ref{fig:curve1} we show how well these
5 models fit the constraints.  As expected from the $\chi^2$ values, the
models reproduce the data with a general accuracy slightly exceeding the 
size of the error bars.  In fact, even these models could be significantly
improved by further local optimizations, because neither the effective
velocity nor the source size is part of the local optimization process
discussed in \S\ref{sec:combine} (only the trajectory starting points and
directions are optimized).  In general, the light curves remain similar
as they interpolate through the gaps in the data, although there is
some divergence for the gap near 900~days. 
This is not true, however, if we extrapolate the behavior over longer time
periods.  Fig.~\ref{fig:curve2} shows the light curves for the same models
but with the time period expanded to cover the 10 years before the OGLE
monitoring period.  For typical models, the source crosses 1--3 Einstein
radii in the OGLE data, so the light curves on longer time periods will
show little correlation with those observed by OGLE.  
Since the OGLE data allows a wide range of magnitude offsets
(\S\ref{sec:fluxrat}, Fig.~\ref{fig:fluxrat}), most of the 
shifts in Fig.~\ref{fig:curve2} are simply due to the difference between
the mean magnification during the OGLE monitoring period and the global
mean. 

Fig.~\ref{fig:srccurve} shows the non-parametric estimates of the intrinsic
source magnitude for the same model realizations.  The offsets in the mean
magnitudes are again due to the lack of convergence to the mean magnification.  
The rms of the intrinsic source
variability ranges from $0.15$~mag to $0.29$~mag, considerably more than the
level of $0.05$~mag we used in our parametric analysis.  The scatter about
a linear trend with time is smaller ($0.12$ to $0.14$~mag).  This suggests
that our parametric models were overly restrictive in their assumptions 
about the source variability, thereby forcing the microlensing variability
to try to model some of the intrinsic variability.  This may explain some
of the velocity shifts between the analyses.  The assumptions of the 
parametric model cannot, however, be completely unrealistic -- for each image 
we do find light curve realizations where a constant source with an rms variability
of $0.05$~mag is statistically consistent with the data.  If such solutions
exist for the individual images, then they also exist for all the images
simultaneously.  They must, however, occupy a small region of the allowed
parameter space.  The source flux variations of our five best solutions are quite
similar (for example, all show a peak near day 1370), so in Fig.~\ref{fig:srccurve}
we also show the statistical average of the source light curves for these
solutions (scaled to the same mean magnitude) and the scatter of the light
curves around the mean.  Despite coming from models in wildly different
regions of the magnification maps (see below), the scatter between the source
light curves is considerably smaller than the overall variations.   This
continues to be true even if we construct the mean source fluctuations including
the next set of 5 best realizations.  Thus, it seems likely that the source
quasar varied by approximately 0.5~mag during the monitoring period with a
peak near day 1400.

\begin{figure}
  INCLUDED ONLY AS JPEG FILE
\caption{ Source trajectories superposed on magnification patterns for the
  best non-parametric realization ($\chi^2=186$ for $N_{dof}=290$).  This
  is a mono-mass, thin disk model with $\log(\hat{r}_s)=15.75$ and $\hat{v}_e=13900$~km/s.  
  The gray scale shows the (unconvolved) magnification pattern for images
  A (top left), B (top right), C (lower left) and D (lower right).  Darker
  colors indicate higher magnifications.  The line shows the source
  trajectory across the pattern for the OGLE monitoring period.  
  The large circle has a radius of
  $\langle\theta_E\rangle$ and the small circle has a radius of 
  $\hat{r}_s$ (the smoothing scale).  The circles are centered on the
  point corresponding to the initial epoch of the OGLE data.
  Depending on the background magnification, the source trajectory
  and the circles are either black or white.
  }
\label{fig:magpat1}
\end{figure}

\begin{figure}
  INCLUDED ONLY AS JPEG FILE
\caption{ Source trajectories superposed on magnification patterns for the
  second best non-parametric realization ($\chi^2=187$ for $N_{dof}=290$).  This
  is a Salpeter, thin disk model with $\log(\hat{r}_s)=15.75$ and
  $\hat{v}_e=14600$~km/s.  See Fig.~{\protect\ref{fig:magpat1}} for 
  description. 
  }
\label{fig:magpat2}
\end{figure}

\begin{figure}
  INCLUDED ONLY AS JPEG FILE
\caption{ Source trajectories superposed on magnification patterns for the
  third best non-parametric realization ($\chi^2=201$ for $N_{dof}=290$).  This
  is a Salpeter, thin disk model with $\log(\hat{r}_s)=15.75$ and
  $\hat{v}_e=29000$~km/s.  See Fig.~{\protect\ref{fig:magpat1}} for  
  description.
  }
\label{fig:magpat3}
\end{figure}

\begin{figure}
  INCLUDED ONLY AS JPEG FILE
\caption{ Source trajectories superposed on magnification patterns for the
  fifth best non-parametric realization ($\chi^2=201$ for $N_{dof}=290$).  This
  is a Salpeter, Gaussian disk model with $\log(\hat{r}_s)=15.50$ and
  $\hat{v}_e=7100$~km/s. 
  See Fig.~{\protect\ref{fig:magpat1}} for  description.
  }
\label{fig:magpat4}
\end{figure}

\begin{figure}
  INCLUDED ONLY AS JPEG FILE
\caption{ Source trajectories superposed on magnification patterns for the
  sixth best non-parametric realization ($\chi^2=201$ for $N_{dof}=290$).  This
  is a Salpeter, thin disk model with $\log(\hat{r}_s)=16.25$ and
  $\hat{v}_e=40200$~km/s.  The trajectory for image B crossed the upper
  edge of the magnification pattern and then continued from the bottom
  edge, which is not shown.
  See Fig.~{\protect\ref{fig:magpat1}} for  description.
  }
\label{fig:magpat5}
\end{figure}

Finally, in Figs.~\ref{fig:magpat1}--\ref{fig:magpat5} we show the source
trajectories generating these light curves superposed on the magnification
patterns.  In order to make the caustics more easily visible, we did not
convolve the patterns with the source structure of the realizations.  
The origin of the scatter in the magnitude offsets $\Delta\mu^\alpha$ 
(\S\ref{sec:fluxrat}) and the offsets in average source brightness 
(Fig.~\ref{fig:srccurve}) are easily understood from these figures.  For
example, image A was used as the magnitude reference point (because
we measured $\Delta\mu^\alpha-\Delta\mu^A$), so the changes in the mean
magnification of image A are responsible for the shifts in the average
magnitude of the source (Fig.~\ref{fig:srccurve}).  In Figs.~\ref{fig:magpat2},
\ref{fig:magpat4} and \ref{fig:magpat5} image A is produced in a 
magnified region, leading to fainter source magnitudes, while in 
Figs.~\ref{fig:magpat1} and \ref{fig:magpat3} it
a demagnified region, leading to a brighter source magnitude.  
 
The magnification patterns are also useful for understanding the origins
of the peaks in the light curves (Fig.~\ref{fig:data2237}).  In particular,
several studies (e.g. Yonehara~\cite{Yonehara01}, Shalyapin et al.~\cite{Shalyapin02}) 
have attempted to model the peaks in the A and C light curves using simple fold caustic 
crossings or isolated point lenses to estimate the source structure.  Sometimes 
models of the peak in the A light curve as a fold crossing is appropriate 
(e.g. Figs.~\ref{fig:magpat1}
and \ref{fig:magpat2}).  But in Figs.~\ref{fig:magpat3}, \ref{fig:magpat4}
and \ref{fig:magpat5} the peak is due to one or more caustic crossings
associated with one or more cusps.  The peaks seen in the light curve of
image C are all associated with cusps, frequently arising from the 
high magnification regions outside the tip of the cusp (e.g. Fig.~\ref{fig:magpat3}).
Wyithe et al.~(\cite{Wyithe00g}) drew a similar conclusion on more
qualitative grounds.
The light curve of image D can be smooth by staying inside the smooth part
of a high magnification region (Fig.~\ref{fig:magpat1}), using the finite
source size to smooth out the variability of a region with very densely
packed caustics (Figs.~\ref{fig:magpat2} and \ref{fig:magpat4}), 
staying in a smooth, demagnified region (Fig.~\ref{fig:magpat3}) or
by putting the caustic crossing inside the monitoring gaps 
(Fig.~\ref{fig:magpat5}).  The shear range of possibilities for producing
quantitatively similar fits does not bode well for attempts to reconstruct
source structures by making simplifying assumptions about the local caustic
structures.

\section{Discussion \label{sec:discuss} }

The method we introduce in this paper reduces the problem of interpreting quasar
microlensing data to a problem of computation rather than conceptualization. Any
quasar microlensing data, from one or more lenses and both more or less complex, 
can be analyzed to derive
physical results.  We demonstrated the method using the most complex, single
quasar microlensing data set, the OGLE light curves for the four images of 
Q2337+0305 to obtain simultaneous constraints on  the microlens mass scale, 
source size, accretion disk structure, and the stellar mass fraction near
the images.  While all these issues have been studied in previous models of
microlensing in Q2237+0305, this is the first time all the relevant physical
properties of the system have been treated simultaneously.  

We estimate that the effective source velocity is fairly high, 
$10200~\hbox{km/s} \ltorder v_e h \avgmhat^{-1/2} \ltorder 39600$~km/s,
which means that the source takes roughly 2 years to move one Einstein
radius.  Because the variability during the OGLE monitoring period was
greater than during most of the preceding decade, the estimate of the
effective velocity may be biased towards higher values than if we had
modeled all the available data.  We estimate statistically that the
source is moving approximately $3300$~km/s from estimates for the peculiar 
velocity of the lens and the velocity dispersion of its constituent
stars.  Combining the probability distributions for the effective
and physical source velocities, we obtain an estimate for the mean
stellar mass of $\langle M \rangle \simeq 0.037 h^2 M_\odot$ 
($0.0059 h^2 M_\odot \ltorder \avgm \ltorder 0.20 h^2 M_\odot$) 
which is somewhat low.  Unfortunately the mass estimate depends
on the square of the velocities, so modest biases in the effective
velocity from using the data during which the variability was largest
or our approximate treatment of the internal motions of the stars
make the systematic uncertainties in the mass estimate difficult to 
evaluate.  Nevertheless, these mass estimates are consistent with previous results 
for this system (e.g. Lewis \& Irwin~\cite{Lewis96},
Wyithe et al.~\cite{Wyithe00b}) and Galactic microlensing studies 
(e.g. Alcock et al.~\cite{Alcock00}). 

The lens galaxy in Q2237+0305 is composed of stars, with a lower bound
of $\kappa_*/\kappa \gtorder 0.5$ on the fraction of the surface mass
density causing the flux variations.  The limit rises to $\kappa_*/\kappa \gtorder 0.7$
if we impose a prior of $0.2 h^2 M_\odot < \avgm < 2 h^2 M_\odot$
on the masses of the microlenses, because models with low $\kappa_*$
require higher effective velocities (Einstein radii per year) corresponding
to lower mass scales $ \avgm $ in order to produce the same amount
of variability.  Since the lensed images in Q2237+0305 are passing
through the bulge of a nearby spiral galaxy (Huchra et al.~\cite{Huchra85})
we expect $\kappa_* \simeq \kappa$ for this system.  However, our ability
to estimate the stellar mass fraction for Q2237+0305 using microlensing
data, indicates that we should also be able to estimate the stellar 
surface density fractions in other lenses where we expect dark matter
to dominate the surface density with $\kappa_*/\kappa \sim 0.1$ to $0.2$
(see Schechter \& Wambsganss~\cite{Schechter02}, 
Rusin, Kochanek \& Keeton~\cite{Rusin03}).
While we kept the properties of the ``macro'' model (the total
surface density and shear for each image) fixed in these calculations,
these parameters could also be constrained by fits to the light curves.
Our models with $\kappa_*/\kappa < 1$ are closely related to models
in which the mass distribution of the lens is more centrally 
concentrated than our standard isothermal model.  This indicates
that the microlensing data will favor the isothermal mass distribution
over more centrally concentrated density profiles.  
  
We find that the data is better fit by a standard thin accretion disk
model than by a Gaussian model of the source's surface brightness. 
We get an accurate estimate of the radius $r_s=2.6_{-1.2}^{+2.0} \times 10^{15}h^{-1}$~cm
at which the disk temperature matches the wavelength of the observations
($2000\AA$ in the rest frame or $T_s\simeq 70000$~K).  The results are
consistent with black body emission and do not require non-thermal or
optically thin emission processes.  We estimate that the black hole
mass is 
$M_{BH}\simeq 1.1_{-0.7}^{+1.4} \times 10^9 h^{-3/2} \eta_{0.1}^{1/2}(L/L_E)^{-1/2}M_\odot$,
which means that $r_s$ corresponds to approximately 8~Schwarzschild
radii from the black hole.  While reassuringly consistent, our treatment
of the source structure has limitations.  First, the physical model
for the accretion disk is more appropriate for the outer regions of 
a thin disk than for the inner regions.  Second, we assumed that the
disk was viewed face-on and was circular.  A more realistic model
would need to use an inclined disk.  

The only practical limitation to our approach is its computational
intensity.  Our present analysis considered 208 different combinations of
stellar density, stellar mass function, source structure and source size,
generating 40~billion non-parametric light curve realizations, 
and required approximately 2~processor-months to do the final calculations.
The problem is, however, trivially parallel, making larger parameter
surveys relatively easy to conduct simply by using more computers (it
would take one day given 60 processors).  Improvements in the sampling
of the variables or the strategies for rapidly discarding poor 
light curve trials should significantly reduce the number of trials
needed to achieve the same statistical results.  For example, we uniformly 
sampled the $\log \hat{r}_s$/$\log\hat{v}_e$ plane, but only a restricted 
region of the plane produces statistically acceptable solutions 
(see Fig.~\ref{fig:rscont}).  One major systematic limitation to
our estimate of the mass scale $\avgm$ is our inability to correctly
treat the internal motions of the stars in the lens galaxy using
static magnification patterns.  Adding the internal motions requires
tracing the source trajectories through a sequence of
magnification patterns (e.g. Wambsganss \& Kundic~\cite{Wambsganss95}).  
This adds little to the execution time,
but requires large amounts of memory.  Models of the OGLE light curves
of Q2237+0305 including the stellar motions require 200-400 time
steps (resolving the mean stellar motion in steps of $0.01$-$0.02\langle\theta_E\rangle$)
for each image, all 13-26~Gbytes of which must be stored in memory.
Fortunately, most multi-processor computers which would significantly
speed the completion of the calculations also have the memory needed
to hold such large data spaces.

At present, only Q2237+0305 has light curve data that justifies such 
computational intensity simply due to the lack of monitoring data for
most lenses.   The Einstein crossing time due to lens motions scales
as  $(1+z_l)(D_{OL}D_{LS}/D_{OS})^{1/2}$, which means that systems 
with low lens redshifts like Q2237+0305 have shorter time scales for
microlensing variability (Kayser \& Refsdal~\cite{Kayser89}). 
 But they are not enormously shorter -- the
other quasar lenses with known redshifts have time scales that are
only 2--3 times longer.\footnote{Although Q2237+0305 has the smallest projected
CMB velocity of the quasar lenses ($57$~km/s), the Einstein crossing
time due to the motion of the observer scales as
$(1+z_l)(D_{OL}D_{OS}/D_{LS})^{1/2}$, which favors low lens redshifts
more strongly than motions due to the lens.  As a result, even the 
lenses with the maximum projected CMB velocity ($370$~km/s) have
crossing times due to our motion only 60\% that of Q2237+0305.} 
Even if the variability rates of the roughly 30 available quasar
lenses are three times slower than in Q2237+0305, monitoring all
of them routinely generates data equivalent to 3 OGLE light curves
each year.  These data can be significantly enhanced by systematically
measuring the differences between the continuum and emission line
flux ratios of the images (e.g. Lewis et al.~\cite{Lewis98}, also
radio, Falco et al.~\cite{Falco96} 
or mid-infrared Agol, Jones \& Blaes~\cite{Agol00}, Wyithe, Agol \&
Fluke~\cite{Wyithe02a}).  Since the emission lines are generated
on scales significantly larger than the continuum, the differences in
the flux ratios provide immediate constraints on the location of the
images in the magnification pattern and on the relative sizes of the
two emitting regions.  A final, but important, advantage of monitoring
as many lenses as possible is that they are statistically independent.
Each new image in a new lens lies in a random region of a new 
magnification pattern, providing new constraints without the long term temporal 
correlations of data obtained by monitoring a particular lens.  Moreover,
estimates of the stellar mass scale $\avgm$ in any particular lens are
ultimately limited by the uncertain peculiar velocity of the lens. Only
by combining the estimates from multiple lenses can we ever obtain
an accurate estimate.

\acknowledgements  I thank R. Di Stefano, D. Rusin, J. Winn \& S. Wyithe for their 
  comments, G. Rybicki \& R. Narayan for discussions about the accretion disk model, 
  L. Hernquist,  K. Nagamine \& V. Springel for computing the rms peculiar 
  velocities of galaxies in the concordance model, and N. Dalal for a copy of his 
  particle-mesh microlensing code which formed the starting point for the subsequent 
  elaborations.    This research was supported by the Smithsonian Institution and NASA 
  ATP grant NAG5-9265.

\appendix

\section{Generating Periodic Magnification Maps \label{sec:rayshoot} }

We use the ray-shooting method (e.g. Schneider et al.~\cite{Schneider92}) to compute
the source plane magnification patterns.  We use a particle-particle/particle-mesh 
(P$^3$M, Hockney \& Eastwood~\cite{Hockney81}) algorithm to separate the long and
short range effects of the stars.  The source plane region is a square with outer
dimension $L_u$, pixel scale $\Delta u$ and a dimension $N_u = L_u/\Delta u$ that
is chosen to be a power of 2.  The image plane is an $L_x \times L_y$ 
rectangle defined by $L_x |1-\kappa-\gamma| = L_y |1-\kappa+\gamma| = L_u$.
The image plane pixel scale is $\Delta x$, so image plane dimensions of
$N_x=L_x/\Delta x$ and $N_y =L_y/\Delta x$ differ.  We choose the larger
dimension of the image plane to be a power of 2.  The smaller dimension of the
image plane is determined by the axis ratio of the rectangle.  In order to
have both square pixels and an exact periodicity of both the source and image 
planes, the integer array dimensions must satisfy 
$N_x |1-\kappa-\gamma| = N_y |1-\kappa+\gamma|$. 
We impose the constraint by first finding the smaller dimension which comes 
closest to satisfying it given the fixed larger dimension and then making
a small adjustment to the shear value ($\Delta\gamma \simeq N_y^{-1} \simeq 0.001$)
so that it becomes exact.  These adjustments are
so small that they have no physical consequences for our results.

The long-range effects of the stars are computed using Fourier methods.  The mass
of each star is assigned to the nearest grid points using  weights determined by
the distance of the star from the pixels (the TSC, triangle-shaped cloud).  We
then compute the deflections produced by the stars by convolving the surface 
density with the deflection kernels ${\bf \alpha_s}$.  We completely separate
the long and short range effects of the gravity using spline models
for the surface density.  We use the spline density distribution which is
\begin{equation}
   \kappa_s(R,s) = { 2 \over \pi s^2 } \left( 1 - { r^2 \over s^2 } \right) 
\end{equation}
for $R < s$ and equal to zero for $R > s$.  For the convolution we use the
deflection pattern of the surface density distribution,
$\kappa_s(R,s)-\kappa_s(R,a)$, which has zero net mass.  The inner scale
$s= 5 \Delta x$ sets the boundary between the the long range and short
range effects of the star.  The outer scale, $a=\hbox{min}(L_x,L_y)/2$, 
guarantees that $\kappa$ is the total surface density.  It does, however,
limit the long range stochasticity of the potential because fluctuations
in the stellar density on scales larger than $a$ are filtered out of the  
gravitational field.
An attentive reader will have noticed that the smaller
dimension of the image plane is not generally a power of 2.  We use the 
FFTW (``Fastest Fourier Transform in the West,'' Frigo \& Johnson~\cite{Frigo98})
Fourier transform package, which is both fast and handles such transforms 
without any special treatment.  This gives the long range deflection field
${\bf \alpha_g}$.
On scales smaller than the inner scale, $s=5\Delta x$, the deflection field
computed from the convolution must be corrected from that of the spline to
that of a real point mass.  Each image plane pixel is associated with a list
of all stars within $s$ of the pixel boundaries.  When we compute ray 
deflections for that pixel we add the true deflection from each of these
stars {\it minus} the contribution from the spline density that we included
in the long range deflection field ${\bf \alpha_g}$ to give the particle
contribution to the deflection ${\bf \alpha_p}({\bf x})$.  

The total deflection is 
\begin{equation}
  {\bf u} = {\bf x} \left( 
                    \begin{array}{cc}
                        1-\kappa-\gamma & 0  \\
                            0           & 1-\kappa+\gamma
                    \end{array} \right)
             - {\bf \alpha_g}({\bf x}) - {\bf \alpha_p}({\bf x}).
  \label{eqn:lens}
\end{equation}
The terms hide two cancellations.  The outer, negative spline density in the gridded
deflection, ${\bf \alpha_g}$, is needed to allow the $\kappa$ in the deflections 
to be the total surface density $\kappa$ rather than $\kappa_s=\kappa-\kappa_*$.  This could be
changed without any particular problem.  The inner spline region ($R< s$) for
each star is added in ${\bf \alpha_g}$ and then subtracted in ${\bf \alpha_p}$
so that the final deflections exactly match that of a point mass.  Note that
a similar scheme would work equally well for models of substructure.  Because
the deflections of the stars are exactly periodic on both the image and
source planes, a single pass over the image plane can identify all rays which
will be mapped onto the source plane and source trajectories can be continuously
traced across the source plane boundaries.  Similarly if we allow the microlenses
to move, their trajectories are periodic on the lens plane grid.
We set all scales using the average Einstein radius of the stars 
$\langle\theta_E\rangle$.  Typically we generated a magnification pattern
with $L_u=40 \langle\theta_E\rangle$ and $N_u=2048$ 
with source plane pixels $\Delta u \simeq 0.02 \langle\theta_E\rangle$.  We
traced rays on a uniform grid with a minimum image plane resolution of 
$0.01 \langle\theta_E\rangle$ and required an average of $100$ rays per 
source pixel.  

\def\bfx{{\bf x}}
\def\bfu{{\bf u}}
Although the magnification patterns for a fixed mass function would appear
to depend on three variables (the smooth  surface density $\kappa$, the stellar
surface density $\kappa_*$, and the shear $\gamma$), the mass sheet degeneracy
(Paczynski~\cite{Paczynski86} for the case of microlensing) means that there
are only two independent variables.  Here we derive a generalized version
of the mass sheet degeneracy.  Consider two systems, labeled A and B, 
defined by point masses with Einstein radii $b_i$ at positions $\bfx_i$
in an external shear $\gamma$, a smooth convergence $\kappa_s$, and a 
mean convergence due to the stars of $\kappa_*$.  The
shear and convergence define a reduced shear $g=\gamma/(1-\kappa_s)$.
The x-component of the lens equations for the two systems are
\begin{equation}
  u_A = (1-\kappa_{sA})(1-g_A)x_A - 
       \sum_i b_{A,i}^2 { x_A - x_{A,i} \over \left| \bfx_A - \bfx_{A,i} \right|^2 }
\end{equation}
and 
\begin{equation}
  u_B = (1-\kappa_{sB})(1-g_B)x_B - 
       \sum_i b_{B,i}^2 { x_B - x_{B,i} \over \left| \bfx_B - \bfx_{B,i} \right|^2 }
\end{equation}
respectively.  Now assume that the two equations can be related by simultaneously
rescaling the source plane coordinates, $\bfu_B=\alpha \bfu_A$, the lens plane
coordinates, $\bfx_B=\beta\bfx_A$, and the Einstein radii, $b^2_{B,i}=\xi b^2_{B,i}$.
For the lens equations this leads to the constraints that the two systems must
have the same reduced shear, $g_A=g_B$, that the convergences are related by
$1-\kappa_{sA} = (1-\kappa_{sB})\beta/\alpha$, and that the Einstein radii are 
related to the coordinate rescalings by $\xi=\alpha\beta$.  The same scalings
hold for the magnifications, with $\mu_B = \beta^2\mu_A/\xi = \beta\mu_A/\alpha$.  
The average 
surface density of the stars transforms as $\kappa_{*B}=\kappa_{*A}\xi/\beta^2$
and the source plane velocity scales as $v_{e,B}=\alpha v_{e,A}$.
The familiar mass sheet degeneracy is found by holding the lens plane scale fixed
($\beta\equiv 1$) and setting $\kappa_{sA}=0$, in which case $\xi=\alpha=1-\kappa_{sB}$,
$b_B^2=(1-\kappa_{sB})b_A^2$, and $\kappa_{*B}=|1-\kappa_{sB}|\kappa_{*A}$.
While there is no new physics in this generalization, it can be computationally
useful.

\end{document}